\documentstyle[twocolumn,eqsecnum,pre,aps]{revtex}
\begin{document}

\twocolumn[\hsize\textwidth\columnwidth\hsize\csname @twocolumnfalse\endcsname
\title{Interface Proliferation and the Growth of Labyrinths in a
Reaction-Diffusion System}
\author{Raymond E. Goldstein\thanks{email: gold@davinci.princeton.edu}}
\address{Department of Physics, Joseph Henry Laboratories,
	Princeton University, Princeton, NJ 08544}
\author{David J. Muraki\thanks{email: muraki@newton.cims.nyu.edu}}
\address{Courant Institute of Mathematical Sciences,
	New York University, New York, NY  10012}
\author{Dean M. Petrich\thanks{email: dpetrich@asphodel.caltech.edu}}
\address{Department of Physics, Joseph Henry Laboratories,
	Princeton University, Princeton, NJ 08544}
\address{$^{**}$Department of Physics, California Institute of Technology,
Pasadena, CA 91125}

\maketitle
\begin{abstract}

In the bistable regime of the FitzHugh-Nagumo model of
reaction-diffusion systems, spatially homogeneous patterns may be
nonlinearly unstable to the formation of compact ``localized states."
The formation of space-filling patterns from instabilities of such
structures is studied in the context of a nonlocal contour dynamics model for
the evolution of boundaries between high and low
concentrations of the activator.  An earlier heuristic derivation
{[}D.M. Petrich and R.E. Goldstein, Phys. Rev. Lett. {\bf 72}, 1120
(1994){]} is made more systematic by an asymptotic analysis
appropriate to the limits of fast inhibition, sharp activator
interfaces and small asymmetry in the bistable minima.  The resulting
contour dynamics is temporally local, with the normal component of the
velocity involving a local contribution linear in the interface
curvature and a nonlocal component having the form of a screened
Biot-Savart interaction. The amplitude of the nonlocal interaction is
set by the activator-inhibitor coupling and controls the ``lateral
inhibition" responsible for the destabilization of localized
structures such as spots and stripes, and the repulsion of nearby
interfaces in the later stages of those instabilities.  The
phenomenology of pattern formation exhibited by the contour dynamics
is consistent with that seen by Lee, McCormick, Ouyang, and Swinney in
experiments on the iodide-ferrocyanide-sulfite reaction in a gel
reactor.  Extensive numerical studies of the underlying partial
differential equations are presented and compared in detail with the
contour dynamics.  The similarity of these phenomena (and their
mathematical description) with those observed in amphiphilic
monolayers, Type-I superconductors in the intermediate state, and
magnetic fluids in Hele-Shaw geometry are emphasized.

\end{abstract}
\vskip2pc]

\section{Introduction}
\label{sec:Intro}

Recent experimental studies \cite{Lee,Lee_long} of pattern formation
in reaction-diffusion systems have revealed a mechanism for the
generation of space-filling patterns that is markedly different from
the classical Turing bifurcation \cite{Turing}.  In the Turing
scenario, realized recently in several experiments
\cite{Boissonade,Swinney,Turing_review}, a periodic pattern arises
throughout space from a linear instability of a homogeneous state.  In
contrast, experiments of Lee, McCormick, Ouyang, and Swinney
\cite{Lee,Lee_long} have shown that space-filling ``labyrinthine"
patterns can develop from finite-amplitude perturbations to linearly
stable homogeneous states (Fig. \ref{expt_fig}).  These labyrinths are
characterized by patches having different chemical composition that
are separated by relatively sharp interfaces, or fronts.  The pattern
formation process involves the motion of these fronts, which are
generally observed to be mutually repelling, thus preventing
self-crossings and associated changes in topology.  The observations
that this system possesses bistability, requires finite-amplitude
disturbances for nucleation of patterns, and displays fingering
instabilities of compact domains suggests a connection to the
one-dimensional, reaction-diffusion ``localized states" considered by
Koga and Kuramoto \cite{Koga}.  Subsequent generalizations to higher
dimensions by Ohta, Mimura, and Kobayashi \cite{Ohta} revealed that
these localized states could exhibit fingering instabilities \cite{Knobloch}.

In earlier work \cite{prl} it was suggested on the basis of heuristic
arguments that this kind of pattern formation by interacting chemical
fronts could be understood by means of a ``nonlocal contour dynamics
model," derived from the well-known FitzHugh-Nagumo model of
activator-inhibitor competition \cite{FitzHugh,Murray} in the limit of
fast inhibition.  This law of motion derives from the combination of a
Young-Laplace force associated with interface curvature \cite{JKeller}
and a screened Biot-Savart coupling between distant segments of the
interface.
This nonlocal contribution embodies the phenomenon of
``lateral inhibition" -- an inhibitory action on a scale much larger
than the activator front thickness \cite{Meinhardt,Ermentrout}.
Numerical studies of the contour dynamics model revealed the essential
features seen in experiments: the instabilities of compact structures,
repulsion of chemical fronts, and the relaxation of branched
structures to compact ones as control parameters are varied.

The competition between Young-Laplace and Biot-Savart forces has been
shown to appear in a variety of other pattern-forming systems, each of
which exhibits labyrinthine interface evolution.  These include
monolayers of dipolar molecules at the air-water interface
\cite{McConnell_Mohwald}, magnetic fluids \cite{Rosensweig} in
Hele-Shaw flow \cite{Cebers,Boudouvis,science}, Type-I superconductors
in the intermediate state \cite{Huebener}, and thin garnet films
\cite{Seul_films}.  In each of these cases, the patterns are defined
by the boundaries between two thermodynamic phases in coexistence,
with bulk electric or magnetic dipolar order.  The Biot-Savart
interactions then arise from the usual correspondence between
magnetization and current loops.  The common phenomenology of these
systems has been reviewed recently \cite{Seul_Andelman}.

Here we elaborate on the contour dynamics in two ways: a systematic
derivation is presented using matched asymptotics, and extensive
numerical evidence is produced to confirm consistency with the
phenomenology of the original reaction/diffusion partial differential
equations (PDEs).  Throughout the analysis we emphasize the
variational structure of the dynamics, both in the fast-inhibitor
limit and more generally, for it appears not to be widely appreciated
that complex
patterns can arise from purely gradient flows.

In order to motivate the major emphases in this work, we show in
Figures~\ref{intro_sim_fig1} and \ref{intro_sim_fig2} two numerical
simulations of the pattern formation exhibited by the FitzHugh-Nagumo
PDE model.  The regions in which the activator takes on each of its
bistable values are indicated by black and white. In
Figure~\ref{intro_sim_fig1} we see a compact initial domain of one
phase expand and finger to produce a space-filling labyrinthine
pattern.  Yet a small change in parameter values leads to the
relaxation of this pattern back to a stable localized state, as shown
in Figure~\ref{intro_sim_fig2} .  In this paper we show that the
complex pattern formation observed in Figures~\ref{intro_sim_fig1} and
\ref{intro_sim_fig2} can in fact be explained in terms of three
elementary processes:

\indent{$\bullet$}  stripe stabilization and repulsion

\indent{$\bullet$}  domain localization

\indent{$\bullet$}  interface proliferation and transverse front instabilities

\noindent The remainder of the paper is organized around the elucidation of
each
of these features on both energetic grounds and by means of asymptotic
methods for the derivation of front motion from PDEs.  The particular
reaction-diffusion model of interest here is introduced in Section II
along with a discussion of its stability and variational structure.
We illustrate the region in parameter space where the Turing
instability is precluded and localized states appear and discuss the
gradient-flow nature of the dynamics in the fast-inhibitor limit.
Interface dynamics in this limit is considered in detail in Sections
III and IV, where we reiterate the heuristic arguments leading to
front dynamics and present the asymptotic analysis.  Section III
focuses on one-dimensional systems where the features of stripe
stabilization (a form of domain localization) and stripe repulsion are
most easily understood.  Two-dimensional systems are considered in
Section IV, in which a variant on domain localization is found.  A
detailed discussion of transverse front instabilities is given in
Sec. V, revealing a common mechanism not unlike that of the
Mullins-Sekerka instability in solidification.  The numerical studies
of both the full PDEs and the contour dynamics are presented in
Section VI.  Some considerations regarding the form of the dynamics
away from the fast-inhibitor limit are outlined in Section VII.
Nonlocal interface dynamics in the various contexts described above
are tied together in Section VIII, and Section IX outlines our
conclusions and open problems.  For completeness, an appendix details
the numerical methods used both for the study of the PDEs and the
contour dynamics.

There have been several other recent theoretical studies relevant to
the experiments of Lee, {\it et al.} In extensive simulations of the
Gray-Scott model \cite{Gray_Scott}, Pearson \cite{Pearson} has found a
wide variety of patterns, including ordered arrays of spots, lamellar
stripe domains, and labyrinthine structures.  Some of these were found
to arise from finite-amplitude perturbations, as in the experimental
work.  More recently, Meron and Hagberg \cite{Meron,Hagberg}, based on
reaction-diffusion equations similar to the FitzHugh-Nagumo model,
have provided elegant geometrical arguments for a connection between
labyrinthine instabilities in the fast-inhibitor limit and spiral wave
behavior for slow inhibition.  The present work is complementary to
these in providing a more detailed picture of the fast-inhibitor
limit, where the dynamics reduces to a pure gradient flow.  Clearly,
however, certain phenomena are precluded by the fast-inhibitor limit,
such as spot ``self-replication" studied in recent experimental
\cite{Lee2} and theoretical works \cite{Pearson,Reynolds,Krischer}.

\section{The Model}
\label{sec:Model}
\subsection{Definition}

Our starting point is the FitzHugh-Nagumo model \cite{FitzHugh} for
the coupled dynamics of an activator $u$ and inhibitor $v$.  We choose
appropriate definitions of $u$, $v$, space and time to arrive at the
nondimensionalized form
\begin{mathletters}
\label{uvequations2}
\begin{equation}
	u_t = D\,\nabla^2 u - F'(u;r) - \rho\,(v-u)
\label{uva2}
\end{equation}
\begin{equation}
	\epsilon v_t = \nabla^2 v - v + u~.
\label{uvb2}
\end{equation}
\end{mathletters}
Here, $D$ is the activator diffusion constant normalized to that of
the inhibitor, and $F(u;r)$ is a double-well potential representing
the autocatalytic behavior of the activator.  In a convenient
parametrization whereby the two local minima in $F$ are at $u=0$ and
$u=1$, the potential $F(u;r)$ is written in terms of a single
symmetry-breaking control parameter $r$ as
\begin{equation}
F(u;r) ={1\over 4}u^2\left(u-1\right)^2
		+\left(r-{1\over 2}\right)
		\left({1\over 2}u^2-{1\over 3}u^3-{1\over 12}\right)
\label{Fdef}
\end{equation}
provided $0<r<1$.  When $r$ differs from $1/2$, a difference in
potential is created between the two states,
\begin{eqnarray}
\Delta F&\equiv& F(u=1;r)-F(u=0;r)\nonumber \\
&=&{1\over 6}\left(r-{1\over 2}\right)
\label{energy_diff}
\end{eqnarray}
so that for $\Delta F>0$ ($r>1/2$) the state $u=0$ is the more stable,
and the reverse for $\Delta F<0$.  In Figures~\ref{intro_sim_fig1}
and~\ref{intro_sim_fig2}, the $u=0$ state is associated with the
{\it white} area and $u=1$ with {\it black}.

With the potential $F$ as defined above, the model PDE is a coupled
diffusion system which is nonlinear only in the activator equation
\begin{mathletters}
\label{uvequations2a}
\begin{equation}
	u_t = D\,\nabla^2u
		-u \left(u-r\right) \left(u-1\right) - \rho\,(v-u)
\label{uva2model}
\end{equation}
\begin{equation}
	\epsilon v_t = \nabla^2 v - \left(v - u\right)~.
\label{uvb2model}
\end{equation}
\end{mathletters}
This particular nondimensional formulation has the advantage of
possessing an {\it exact} invariance under the simultaneous
transformations
\begin{equation}
	u\to 1-u~; \ \ \ \ v\to 1-v~; \ \ \ \ r\to 1-r~.
\label{symmetries}
\end{equation}
This property implies that we need not consider as distinct cases the
evolution of black spots in white domains versus white spots in black
domains.  Within the context of the PDE model (\ref{uvequations2a}) we
therefore focus on the question:
\vskip 0.3cm

\indent ``By what mechanism do black labyrinthine patterns develop in a
white domain?"

\vskip 0.3cm

The form of the inhibitor dynamics embodies both its self-limiting
behavior and its stimulation by the activator, while the linear
coupling to $v$ in the activator dynamics is the simplest such term.
Although it is traditional to view $v$ itself as the inhibitory agent,
we have chosen instead to introduce an inhibitor coupling in the form
$\rho(v-u)$ which insures that the state $u=v=1$ remains a stationary
uniform state for all $\rho$.

The parameter $\epsilon$ in Eq. \ref{uvb2} distinguishes between the
slow-inhibitor limit ($\epsilon \gg 1$) and the fast-inhibitor regime
($\epsilon \ll 1$).  We assume the latter, a limit that is {\it
opposite} to the limit assumed in phase-field models \cite{Fife} and
spiral wave dynamics \cite{spiral_waves}.  Through the
nondimensionalizations used in (\ref{uvequations2a}), the natural
diffusive length scale of the inhibitor is set to unity.  The
corresponding length scale for the activator is $\sqrt{D}$, so the
limit of sharp activator fronts of interest here requires $\sqrt{D}\ll
1$.  The parameter $D$ serves to mainly to set the width of those
fronts, while the two remaining parameters in the model, $r$ and
$\rho$, will be seen to control more global aspects of
the front dynamics.

\subsection{Stability against the Turing bifurcation}

A simple stability analysis shows that both states $u=0$ and $u=1$
can be linearly stable {\it simultaneously} to all periodic disturbances.
Consider first the state $u=v=0$.  In the limit of fast inhibition,
$\epsilon \to 0$ and for $0<r<1$ (so that both minima of $F$ exist),
one finds that perturbations having the form
\begin{eqnarray}
u&=&U{\rm e}^{ikx+\sigma t}\nonumber \\
v&=&V{\rm e}^{ikx+\sigma t}
\label{perts}
\end{eqnarray}
are characterized by two branches of solutions for the growth rates $\sigma$,
\begin{eqnarray}
\sigma_+&=& -Dk^2 -r+\rho{k^2\over 1+k^2}
+ {\cal O}(\epsilon)\nonumber \\
\sigma_-&=& -{1+k^2\over \epsilon} + {\cal O}(1)~.
\label{growths}
\end{eqnarray}
While the branch $\sigma_-$ is clearly damped for all wavevectors,
$\sigma_+$ may become positive for $\rho$ exceeding a critical value
$\rho_c(r)$.  The neutral curve for this linear (Turing) instability, as
defined by
the conditions $\sigma_+(k)=0$ and $d\sigma_+(k)/dk=0$, is
\begin{equation}
\rho_T(r) \sim \left(\sqrt{r}+\sqrt{D}\right)^2
\label{neutral}
\end{equation}
so that when $\rho < \rho_T(r)$ the state $u=v=0$ is linearly stable.
At the critical $\rho=\rho_T(r)$ the marginally stable wavevector
$k_T$ is
\begin{equation}
k_T \sim \left(\sqrt{{\rho_T\over D}}-1\right)^{1/2}
		= \left({r\over D}\right)^{1/4}~.
\label{neutral_k}
\end{equation}

Appealing to the symmetry relations in Eq. \ref{symmetries}, we deduce
that the stability criterion for the state $u=v=1$ is
$\rho < \rho_T(1-r)$, where the function $\rho_T$ the same as in
(\ref{neutral}).  Thus, {\it simultaneous} stability of both
homogeneous states to the Turing bifurcation requires $\rho<\rho_T(r)$
{\it and} $\rho<\rho_T(1-r)$.  The region in $\rho-r$ space so defined
is shown in Figure \ref{linstab_fig} for the case $D=0.01$.  The inset
to Fig. \ref{linstab_fig} shows two curves of $\sigma_+(k)$ for the
parameters ($r=0.65,\rho=0.25$) (stable) and ($r=0.65,\rho=0.60$)
(unstable).

As the markers on Figure~\ref{linstab_fig} suggest, the labyrinthine
dynamics of Figures~\ref{intro_sim_fig1} and~\ref{intro_sim_fig2} are
observed near $\rho~\approx~0$ and $r~\approx~1/2$ -- values which are
well within the stable regime.  Although for these parameters the
system is stable against the Turing mechanism, it is the underlying
bistability that permits the existence of non-trivial localized states
through intrinsically nonlinear processes.

\subsection{Variational Aspects}

Equations \ref{uvequations2a} are neither purely dissipative nor
Hamiltonian, but can be expressed in a variational form
\begin{mathletters}
\label{variational}
\begin{equation}
u_t=-{\delta {\cal E}_u\over \delta u}-\rho{\delta {\cal F}\over \delta u}~,
\label{uvar}
\end{equation}
\begin{equation}
\epsilon v_t=- {\delta {\cal E}_v\over \delta v}
+{\delta {\cal F}\over \delta v}~,
\label{vvar}
\end{equation}
\end{mathletters}
where the energy functionals are
\begin{eqnarray}
{\cal E}_u &=& {\int\! d{\bf x}}
	\left\{{1\over 2} D \vert\bbox{\nabla} u\vert^2
		+F(u;r)-{1\over 2}\rho u^2\right\}\nonumber \\
{\cal E}_v &=& {\int\! d{\bf x}}
	\left\{{1\over 2}\vert\bbox{\nabla} v\vert^2
		+{1\over 2} v^2\right\}\nonumber \\
{\cal F} &=& {\int\! d{\bf x}}\ uv~.
\label{energy_func}
\end{eqnarray}
It is precisely because the cross-terms in the dynamics
(\ref{uvequations2a}) are of opposite sign that the system is not a
pure gradient flow for finite $\epsilon$.  One may verify that neither
${\cal E}_u$, nor ${\cal E}_v$, nor ${\cal F}$ decreases monotonically
in time.  This completely changes in the fast-inhibitor limit
\cite{Bernoff_private} ($\epsilon \to 0$), for then Eq. \ref{vvar}
becomes a functional relation
\begin{equation}
{\delta {\cal E}_v\over \delta v}={\delta {\cal F}\over \delta v}
\label{func_rel}
\end{equation}
from which it follows that
the combination ${\cal E}_u+\rho {\cal F}$ decreases monotonically in time
\begin{equation}
{\partial \over \partial t}\left({\cal E}_u+\rho{\cal F}\right) =
	-{\int\! d^2x}\left({\delta {\cal E}\over \delta u}
		+\rho {\delta {\cal F}\over \delta u}\right)^2~.
\label{yup}
\end{equation}
The dynamics is then a gradient flow, a feature central to the
heuristic arguments given in earlier work \cite{prl} and elaborated on
below.

\subsection{The Nonlocal Energy Functional}

With the exception of Section VII, for the remainder of this paper we shall
assume the fast-inhibitor limit $\epsilon=0$.  The inhibitor dynamics
degenerates into an instantaneous-in-time relation between $v$ and $u$
given by Eq. \ref{func_rel}.  Using the specific forms of ${\cal E}_v$
and ${\cal H}$ it reduces to
\begin{equation}
\left(\nabla^2-1\right)	v({\bf x},t) = -u({\bf x},t)
\label{fast_inhibitor}
\end{equation}
which can be solved using a Green's function
\begin{equation}
v({\bf x},t)=\int\! d{\bf x}'{\cal G}({\bf x}-{\bf x}')u({\bf x}')~.
\label{slave}
\end{equation}
In one and two dimensions respectively, the Green's functions are
\begin{mathletters}
\label{Green_define}
\begin{equation}
{\cal G}(x-x')={1\over 2}{\rm e}^{-\vert x-x'\vert}
\label{1dGreen}
\end{equation}
\begin{equation}
{\cal G}({\bf x}-{\bf x}')={1\over 2\pi}K_0(\vert {\bf x}-{\bf x}'\vert)
\label{2dGreen}
\end{equation}
\end{mathletters}
where $K_0$ is the modified Bessel function of order zero.

Substituting for $v$ in Eq. \ref{uva2model}, we obtain the spatially
nonlocal activator PDE
\begin{equation}
u_t=D\nabla^2u-F'(u;r)+\rho u
-\rho\int\! d{\bf x}'{\cal G}({\bf x}-{\bf x}')u({\bf x}')~.
\label{slave1}
\end{equation}
{}From the arguments leading up to Eq. \ref{yup}, and as had been
observed earlier by Ohta, Ito, and Tetsuka \cite{Ohta2},
Eq. \ref{slave1} has the variational form $u_t=-\delta {\cal E}/
\delta u$ with the energy functional
\begin{eqnarray}
{\cal E}[u]&=&\int\! d{\bf x}
\left\{{1\over 2} D \vert\bbox{\nabla} u\vert^2
+ F(u;r)-{1\over 2}\rho u^2\right\}\nonumber \\
&&+{1\over 2}\rho \int\! d{\bf x}\!\int\! d{\bf x}'
u({\bf x}){\cal G}({\bf x}-{\bf x}') u({\bf x}')~.
\label{slaved_energy}
\end{eqnarray}
While, as remarked earlier, ${\cal E}$ decreases monotonically in
time, ${\cal E}$ is not necessarily bounded from below in an infinite
domain, and this variational principle does not guarantee a stationary
long-time limit.

As an aside, if the variations in $u$ are on length scales long
compared to the ${\cal O}(1)$ inhibitor screening length, then we can
expand the integrand in (\ref{slaved_energy}) for ${\bf x}'$ near
${\bf x}$ and obtain up
to fourth order the gradient expansion
\begin{equation}
\int\! \! \int\! u {\cal G}u\simeq \int\! d{\bf x}\left(u^2-\vert\bbox{\nabla}
u\vert^2
+\left(\nabla^2 u\right)^2 +\cdots\right)~.
\label{slaved_energy_local}
\end{equation}
Collecting together terms, we obtain an approximate local theory,
\begin{equation}
{\cal E}\simeq\int\! d{\bf x}\left\{{1\over 2} {\bar D} \vert\bbox{\nabla}
u\vert^2 +
{1\over 2}\rho\left(\nabla^2 u\right)^2 +F(u;r)\right\}
\label{local_approx}
\end{equation}
where ${\bar D} = D-\rho$.  At this level, the energy functional is
very similar to that which appears in the Swift-Hohenberg model
\cite{Swift_Hohenberg}, as well as the theory of Lifshitz points in
condensed matter systems \cite{Lifshitz}, and leads quite naturally to
modulated patterns by virtue of the possibly negative coefficient of
$\vert\bbox{\nabla} u\vert^2$.  We note however, that such a gradient
expansion is only of limited utility in the limit of sharp interfaces
of interest in the present work.  Nevertheless, an approximate local
theory can be developed within the limit of sharp interfaces when
the curvature of the interface is small, as shown in Section V.B.

Let us note finally one important feature of
Eq. \ref{local_approx}.  We see that the lowest-order term in the
gradient expansion (\ref{slaved_energy_local}) leads to a contribution
$(1/2)\rho u^2$ in the energy functional.  This precisely cancels the
term $-(1/2)\rho u^2$ seen in Eq. \ref{slaved_energy} which arises
from the $-\rho(v-u)$ inhibition coupling. Thus, the effective
potential in a local theory (\ref{local_approx}) is precisely the
function $F(u;r)$, a convenient simplification in subsequent sections.

\section{Interface Motion in One Dimension}
\label{sec:Onedim}
\subsection{Energetic Derivation of a Stripe Evolution}

In the PDE dynamics shown in Figs. \ref{intro_sim_fig1} and
\ref{intro_sim_fig2}, the potential difference $\Delta F$ is positive
so that $u=0$ (white) is the energetically-preferred state.
Nonetheless, in both cases the black domain survives despite its
energetic disadvantage.  Insight into this bistable configuration can
be obtained through an analysis of a single black stripe.

Consider the dynamics of an interval with $u \approx 1$ in a
background of $u \approx 0$ (i.e. the cross-section of a black
stripe).  A PDE computation of (\ref{uvequations2a}) for one such
evolution is shown in Fig. \ref{localized_state_fig}, where we see an
approach of the two {\it sharp} activator $(u)$ fronts that mark the
transition from black to white.  In contrast, the inhibitor field $v$
is considerably more diffuse.  But, perhaps the most notable aspect
about this particular evolution is that the black stripe
asymptotically stabilizes at a finite size.  By applying a simple
heuristic argument that was outlined in an earlier work \cite{prl}, an
approximate dynamics for this process can be derived which predicts
the equilibrium width of this black ``localized state."

The argument begins from the essential assumption that the activator
fronts are narrow relative to the scale of the pattern, so that $u$
may be taken to be piecewise constant away from fronts.  We focus on
the case of a single black stripe of $u=1$ located symmetrically
between $x=-Q(t)$ and $x=+Q(t)$ in an otherwise quiescent white
background ($u=0$).  The evolution of $Q(t)$ is deduced from
the gradient dynamics based on the one-dimensional version of the
energy functional (\ref{slaved_energy})
\begin{eqnarray}
{\cal E}[u]&=&\int_{-\infty}^{\infty}\! dx \left\{{1\over 2}
Du_x^2 + F(u;r)-{1\over 2}\rho u^2\right\}\nonumber \\
&&+ {1\over 2}\rho\int_{-\infty}^{\infty}\!
dx \int_{-\infty}^{\infty}\! dx'u(x){\cal G}(x-x')u(x')~.
\label{E_one_dimension}
\end{eqnarray}
To insure a finite energy, we consider the energy difference $\Delta
{\cal E}={\cal E}[u]-{\cal E}[u_{\infty}]$, where $u_{\infty}$ is the
value $u$ takes as $x\to \pm \infty$.  Then, in following the
prescription of the earlier results \cite{prl}, the energy functional
for a stripe is partitioned into three contributions: an effective
line tension $\gamma$, a pressure $\Pi$, and a nonlocal coupling
between fronts.  These contributions are explicitly calculated below.

In calculating $\Delta {\cal E}$ for the stripe, the line
tension consists of the contributions that are localized at the fronts
(like $u_x^2$).  In one dimension, this requires a contribution to
$\Delta {\cal E}$ that is some constant $\gamma$ proportional to the
number of fronts.  On dimensional grounds, the line tension associated
with a single front $\gamma$ can be estimated by the gradient integral
\begin{equation}
\gamma \approx \int_{\rm front}\! dx\,D\,u_x^2 =
{\cal O}\left({D\over \ell}\right)~,
\label{gamma_define}
\end{equation}
where $\ell$ is a characteristic length scale of the front.  Since
this length scale must be the diffusive length $\sqrt{D}$, we expect
$\gamma\sim\sqrt{D}$.

Under the assumption that $u$ is piecewise constant and either zero or
unity, the evaluation of the pressure and nonlocal contributions to
the energy functional are considerably simplified.  By virtue of
having subtracted the white background energy ${\cal E}[u_{\infty}]$,
the contribution arising from the potential $F$ is restricted to the
stripe, and is thus proportional to the pulse length $2Q$.  The
constant of proportionality associated with the integral of the
potential function $F$ is interpreted as a pressure $\Pi$ and is given
by
\begin{equation}
\Pi \equiv {1\over 2Q} \int_{-Q}^{+Q}\! dx
\left\{F(1;r)-F(0;r)\right\} = \Delta F~.
\label{energy_est}
\end{equation}

The remaining two terms in the energy integral (\ref{E_one_dimension})
are proportional to the inhibition coupling $\rho$ and are also
integrations only over domains where $u=1$.  For the single black
stripe, these two integrals, when combined, require the evaluation of
\begin{equation}
-{1\over 2}\rho\int_{-Q}^{Q}\! dx
\left\{1 - \int_{-Q}^{Q}\! dx'{\cal G}(x-x')\right\}~.
\label{details1}
\end{equation}
Using the PDE for the Green's function ${\cal G}={\cal
G}_{xx}+\delta(x-x')$ and the reciprocity relation ${\cal
G}_{xx}=-{\cal G}_{xx'}$, the integral (\ref{details1}) simplifies to
\begin{equation}
-{1\over 2}\rho\int_{-Q}^Q\! dx \int_{-Q}^Q\! dx'
{\cal G}_{xx'}= \rho \left[{\cal G}(2Q)-{\cal G}(0)\right]~.
\label{energy_est2}
\end{equation}

Combining all three contributions
(\ref{gamma_define}, \ref{energy_est}, \ref{energy_est2})
the energy of a localized state of size $2Q$ is thus estimated to be
\begin{equation}
\Delta{\cal E}[Q] \simeq 2\gamma + 2Q\,\Pi + \rho\left[{\cal G}(2Q)
-{\cal G}(0)\right]~.
\label{energy_est3}
\end{equation}
As remarked in the paragraph following Eq. \ref{local_approx}, we see
that the pressure contribution $\Pi$ comes from the bare function
$F(u;r)$, rather than from the shifted function $F-(1/2)\,\rho u^2$ in
Eq. \ref{slaved_energy}.  Furthermore, the line tension constant
$\gamma$ as estimated from Eq. \ref{gamma_define} is in fact
asymptotically correct, despite an apparent discrepancy of a factor of
two between (\ref{E_one_dimension}) and (\ref{gamma_define}).  This
detail is resolved by the asymptotic analysis in the following
section, and involves the recognition of an additional ${\cal
O}(\sqrt{D})$ contribution from a boundary correction to the estimate
in (\ref{energy_est}).

The final step in this heuristic derivation is the translation of the
variational principle $u_t=-\delta {\cal E}/\delta u$ into an equation
of motion for the front position $Q(t)$.  In terms of the general
theory of Lagrangian dynamics, viscous forces can be introduced as the
variational of a Rayleigh dissipation functional ${\cal R}$
\cite{Goldstein},
\begin{equation}
{\cal R}[u_t]\equiv{1 \over 2}~\int_{-\infty}^{+\infty}\! dx \,u_t^2
\label{gen_disspn_fnc_1d}
\end{equation}
so that the dynamics can be written in the variational form
\begin{equation}
{d\over dt}{\partial{\cal L}\over \partial u_t} -
{\partial {\cal L}\over \partial u} =
-{\partial {\cal R}\over \partial u_t}~.
\label{gen_Lagrang}
\end{equation}
For a single moving front having an invariant spatial profile, the
dissipation functional is intimately related to the line tension
constant $\gamma$ since
\begin{eqnarray}
{\cal R}[u_t] &\equiv & Q_t^2\int_{\rm front}\!dx\,u_x^2 \nonumber \\
& \approx &{\gamma \over D}Q_t^2~.
\label{disspn_fnc_1d}
\end{eqnarray}
where the result is doubled to account for both fronts.  Now, adopting
$Q$ as the dynamical coordinate and taking for the Lagrangian
the energy difference ${\cal L}[Q] = -\Delta{\cal E}[Q]$, the
variational principle (\ref{gen_Lagrang}) gives a dynamical equation
for the front location $Q(t)$
\begin{equation}
Q_t=-{D \over 2\gamma}{\partial \Delta{\cal E}\over \partial Q}=
-{D \over \gamma}\left[
\Delta F - {\rho \over 2}{\rm e}^{-2Q}\right]~.
\label{Q_dynamics_heuristic}
\end{equation}
Note that the absence of any $u_t$-dependence in the Lagrangian ${\cal
L}[u]$ leads to an inertialess front dynamics. The possibility of
inertial contributions to the front evolution is discussed Section VII.

We deduce from (\ref{Q_dynamics_heuristic}) that black localized
states are only possible if $\Delta F>0$, so that the state $u=1$ has
the higher energy of the two minima of $F(u)$.
Moreover, beyond the threshold $\rho=2\Delta F$ the inhibitor coupling
supports a stable black stripe of width
\begin{equation}
2Q^*\sim \ln\left({3\tilde\rho\over \tilde r}\right)~,
\label{L_equil}
\end{equation}
and whose associated energy is
\begin{equation}
{2\Delta {\cal E}\over \sqrt{D}}=
{\sqrt{2}\over 3}-\tilde\rho\left\{1-{\tilde r\over 3\tilde \rho}
\left[1-\ln\left({\tilde r\over 3\tilde \rho}\right)\right]\right\}~.
\label{delta_E_nifty}
\end{equation}
We have introduced the scaled parameters
\begin{equation}
\tilde r={6\Delta F\over \sqrt{D}}={r-1/2\over \sqrt{D}}~, \ \ \ \ \tilde
\rho={\rho\over \sqrt{D}}~,
\label{rescaled_rho_r}
\end{equation}
which in the following section are shown to follow naturally from an
asymptotic analysis.
Below threshold, the inhibition coupling is insufficient to prevent
annihilation through the collision of two fronts.  On the other hand,
for $\Delta
F<0$, no equilibrium width exists and the stripe expands indefinitely
to fill space.  The existence of an equilibrium stripe width for the
{\it less-preferred} state demonstrates, within the context of this
PDE model, the process of {\it stripe stabilization}.

Using the symmetry relations (\ref{symmetries}), we deduce the
converse result that white localized states ($u=0$) in a black
background ($u=1$) may exist only when $\Delta F<0$.  Likewise, a
white stripe will expand unboundedly when $\Delta F>0$.

Localized states may also be formed by the nucleation of a small
domain of $u=1$, followed by its growth to the stable size. From the
form of the energy (\ref{energy_est3}) we see that there is a barrier
to the creation of a localized state, so this process requires a
nucleus larger than some critical size.  This is illustrated by the
PDE computation in Fig. \ref{localized_state_fig2}, where we see the
growth of a slightly supercritical nucleus.

While we have emphasized the
somewhat counterintuitive result that a
localized {\it one-dimensional} state consists of the phase which has a
higher value of the potential $F$, it is interesting to note
from (\ref{delta_E_nifty}) that
the total ``structural energy" $\Delta {\cal E}$ of the stripe
can be of either sign.
It will ultimately be shown (Sec. \ref{sec:Instab}) that the
existence of stripes with $\Delta {\cal E}=0$ is intimately connected with
the onset of the labyrinthine instability.

\subsection{Asymptotic Derivation of a Front Dynamics}

The previous section describes the dynamics and interaction of
interfacial fronts as a Lagrangian relaxation process which is
independent of the detailed fine structure of the activator ($u$)
field.  Using formal matched-asymptotic arguments
\cite{JKeller,Fife,spiral_waves,Type_Ia}, these results are substantiated
as the leading order behavior of the PDE model \ref{uvequations2} in a
particular limit of weak activator diffusion.

The asymptotic analysis relies on a strong separation of scales
between the spatial structures of the activator and inhibitor fields.
Through the non-dimensionalization of the reaction/diffusion PDE
\ref{uvequations2}, the $O(1)$ spatial scale is defined to be the
natural diffusion length of the inhibitor ($v$) field.  In the limit
of weak activator diffusion ($D \ll 1$) the transitions of the
activator field between bistable states are resolved within thin
interfaces, or interior layers, whose $O(\sqrt{D})$ width is
characteristic of diffusion-limited fronts.  From the perspective of
matched-asymptotics, the sharp activator fronts on the {\it inner}
scale are contrasted by the more gradual relaxation of the inhibitor
concentration that occurs on the {\it outer} scale.

{\it Outer $O(1)$ Scale}:  In one dimension ($x$), the model equations
\ref{uvequations2} are rewritten explicitly
\begin{mathletters}
\label{uvequations3}
\begin{equation}
u(u-r)(u-1) = D u_{xx} - \rho (v-u) - u_t~,
\label{uva3}
\end{equation}
\begin{equation}
v_{xx} - (v-u) = \epsilon v_t~.
\label{uvb3}
\end{equation}
\end{mathletters}
where the activator diffusion ($D \ll 1$) and inhibition coupling
($\rho \ll 1)$ are assumed small.  In anticipation of a dynamics
consisting of slow front motion (implied by $D \ll 1$), both
time-derivatives may also be consistently assumed small.  The
right-hand sides of \ref{uvequations3} are then seen to be
perturbations with respect to this outer scale.

In terms of the outer variables, the underlying pattern is defined by
regions where the activator is in either of its locally stable values.
For the cubic nonlinearity (\ref{uva3}), these are
\begin{equation}
u(x) \sim u^0(x) = \cases{
1&$\mbox{black areas}$~\cr
0&$\mbox{white areas}$\cr}
\label{u_outer}
\end{equation}
which defines a binary pattern on the line.  The present approach is more
general than the heuristic analysis of the previous section, capable of
treating a pattern consisting of any number of fronts.  The leading order outer
inhibition field, $v^0(x)$, then satisfies the inhomogeneous elliptic
equation
\begin{equation}
v^0_{xx} - v^0 \sim - u^0 =
-\cases{1&$\mbox{black areas}$~\cr
0&$\mbox{white areas}$~,\cr}
\label{outer_v}
\end{equation}
where continuity of $v^0$ and $v^0_x$ is imposed at jump
discontinuities in the binary pattern $u^0(x)$.  As well, suitable
boundary conditions at the domain edges (for example: periodic, or
$v_x=0$ at infinity) must be included to insure a unique solution.

The time dependence of this leading order outer solution is implicit
in the functions $Q^{\,j}(t)$ that locate the boundary points between the
black and white areas, the jump transitions of $u^0$.
The velocity $Q^{\,j}_t$ of these interface points is obtained as a
consequence of the asymptotic matching at inner scales to remove the
discontinuities of the chemical fields.

{\it Inner $O(\sqrt{D})$ Scale}: At each interface point $x=Q^{\,j}(t)$
the discontinuity of the outer activator solution
$u^0(x)$ is resolved by an inner representation defined locally on a
finer diffusion scale $\eta=(x-Q^{\,j})/\sqrt{D}$.  Since the activator
solution is not discontinuous on this inner scale, the precise
location of the interface $(\eta=0)$ is defined by the condition
$u(x=Q^{\,j})=1/2$.  In terms of this rescaled and moving coordinate, the
equations for the inner solutions $U(\eta,t)$ and $V(\eta,t)$ reflect
a markedly different asymptotic ordering.  Introducing the nonlinear
operator ${\cal S}[U]$
\begin{equation}
{\cal S}[U]=U_{\eta\eta}-U~(U-1/2)~(U-1),
\label{L_define}
\end{equation}
the inner equations become
\begin{mathletters}
\label{inner_equations}
\begin{eqnarray}
{\cal S}[U]&=&-{Q^{\,j}_t\over \sqrt{D}}U_{\eta}
-6\Delta FU\left(U-1\right)\nonumber \\
&&+\rho\left(V-U\right)+U_t
\label{innera}
\end{eqnarray}
\begin{equation}
V_{\eta\eta} = D\left(V-U\right)+\epsilon DV_t~.
\label{innerb}
\end{equation}
\end{mathletters}
The homogeneous solution of the left-hand sides of (\ref{inner_equations})
correspond to a stationary front solution
\begin{eqnarray}
	U^0(\eta) &=& {1\over 2} \left[1
		\mp\tanh\left({\eta\over 2 \sqrt{2}}\right)\right]~,
			\nonumber \\
	V^0 &=& v^{0}(x=Q^{\,j})~,
\label{hyp_tan}
\end{eqnarray}
which, as $\eta \rightarrow \pm \infty$, match to the behaviors of the
outer solutions $u^0(x)$ and $v^0(x)$ at the interface $x\rightarrow
(Q^{\,j})^{\pm}$.  The sign $\mp$ is chosen by the orientation of the
front, where a negative sign corresponds to a left-to-right transition
from $u=1$ to $u=0$.  Note also that, to leading order, the inhibitor
field is constant within this interior layer -- this is consistent
with the continuity of the outer solution $v^0(x)$ of (\ref{outer_v})
which only requires an $O(D)$ transition solution to resolve the
discontinuity in its second derivative.

The stationarity of this leading order front (\ref{hyp_tan}) is broken
by the effects of the right-side terms of (\ref{inner_equations}) and
suggests a perturbative derivation for the front speed $Q^{\,j}_t$ based
on the asymptotic balance
\begin{equation}
{Q^{\,j}_t \over \sqrt{D}} \sim \Delta F \sim  \rho \ll 1~.
\label{pert}
\end{equation}
The determination of $Q^{\,j}_t$, which is essentially just a nonlinear
eigenvalue for (\ref{inner_equations}), then follows from the
solvability condition for a (bounded) first correction to the inner
activator solution
\begin{equation}
U(\eta) \sim U^0(\eta)+\rho U^1(\eta)~.
\label{u_inner}
\end{equation}
Substitution into (\ref{innera}) gives an inhomogeneous ODE for the
time-independent first correction $U^1(\eta)$ involving the
operator ${\cal S}^\prime[U^1]$ that is just the linearization of
the nonlocal operator ${\cal S}[U]$ (\ref{L_define})
\begin{equation}
{\cal S}'[U^1]=U^1_{\eta\eta}
- F^{\prime\prime}\!\left(U^0;1/2\right)U^1
\label{Lprime_define}
\end{equation}
where the primes on $F$ denote differentiation on the first variable.
The first-order perturbations are restricted to the terms involving
the balance (\ref{pert}) and gives
\begin{eqnarray}
\rho {\cal S}^\prime[U^1]&=&
\left[\pm{Q^{\,j}_t\over\sqrt{2D}}
+ 6\Delta F\right]~U^0\left(U^0-1\right)\nonumber \\
&&+ \rho \left[v^0(x=Q^{\,j})- U^0\right]~.
\label{f_corr}
\end{eqnarray}
Application of the identity $\sqrt{2}\,U^0_\eta = \mp U^0\,(U^0-1)$
has resulted in some simplification of terms in the above equation.

By the translation symmetry associated with the $U^0$-solution, the
appropriate solvability condition for bounded solutions to
(\ref{f_corr}) is the inner product integral
\begin{equation}
\langle U^0_\eta,\bullet\rangle \equiv
\int_{-\infty}^{+\infty}\!\!d\eta \,U^0_\eta\bullet ~.
\label{solve}
\end{equation}
Since the linearized operator ${\cal S}^\prime$ is orthogonal to
$U^0$, the inner product of the equation (\ref{f_corr}) determines the
unique front speed that ensures the existence of a bounded first
correction $U^1(\eta)$
\begin{equation}
Q^{\,j}_t \sim \mp 6\sqrt{2D}\left\{\Delta F
+ \rho \left[v^0\left(x=Q^{\,j}\right) - {1\over 2}\right] \right\}~,
\label{speed}
\end{equation}
where again the choice of sign $\mp$ is consistent with the choice
made for the leading-order solution (\ref{hyp_tan}).
Self-consistently, the front velocity satisfies $Q^{\,j}_t \ll {\cal
O}\sqrt{D}$.

The important feature of the front speed result (\ref{speed}) is its
dependence on the strength of the outer inhibition $v^0(x=Q^{\,j})$.
By the elliptic nature of the outer inhibitor equation
(\ref{outer_v}), $v(x=Q^{\,j})$ will in general be influenced by the
location of other fronts -- and hence embody the nonlocality.  Thus
for a sequence of fronts, the speed relations (\ref{speed}) for $j=1$
to $N$ will be a set of $N$ coupled ODEs that describe the boundaries
of the binary pattern for $u(x)$.

{\it Motion of a single front.} For this case, on the infinite line,
we consider a pattern that is white ($u=0$) to the right of an
interface at $x=Q(t)$ and black ($u=1$) to the left,
\begin{equation}
u^0(x) \sim \cases{
1&$\mbox{}x \le Q(t)$~\cr
0&$\mbox{} Q(t) \le x$~.\cr}
\label{one_fr_u}
\end{equation}
Solving (\ref{outer_v}), the outer inhibitor field has the simple
piecewise continuous solution
\begin{equation}
v^0(x) \sim \cases{
1 - \frac{1}{2} {\rm e}^{+(x-Q)}& $\mbox{   } x \le Q(t)$~\cr
\frac{1}{2} {\rm e}^{-(x-Q)}&$\mbox{   } Q(t) \le x$~,\cr}
\label{one_front_v}
\end{equation}
where $v_x=0$ boundary conditions are imposed at infinity.  Since
$v^0(x=Q) = 1/2$ at the interface, by (\ref{speed}) there is seen to
be no $O(\rho)$ correction to the front speed
\begin{equation}
Q_t \sim - 6\sqrt{2D}~\Delta F~,
\label{one_front_Q}
\end{equation}
so that, when $\Delta F >0$, the front travels with constant negative
(leftward) velocity.  Motion towards the black region is the expected
result when the white $(u=0)$ state is energetically-preferred
(\ref{Fdef}).

{\it Higher-order corrections to the speed of a single front.}
The calculation of the front profile and speed may be
continued to another order in $\rho$ in the case of a solitary front.
The solution of (\ref{f_corr}) for the first correction $U^1$ is
\begin{equation}
U^1(\eta) = - {d\over d \eta}
\left\{\eta\tanh\left({\eta \over 2 \sqrt{2}}\right)\right\}~.
\label{another?}
\end{equation}
This describes the overshoot of the profile seen in Figs.
\ref{localized_state_fig}
and \ref{localized_state_fig2}.
The gradient of the front
is increased in the neighborhood of $\eta=0$, and this would be
expected to increase the rate of dissipation, and hence retard the
interface motion.  This intuition is confirmed asymptotically when
$\Delta F \sim \rho \gg \sqrt{D}$, as the next order of the solvability
condition can
be explicitly evaluated, giving the extended velocity condition
\begin{equation}
Q_t \sim - 6\sqrt{2D}~\Delta F~\left[1 - 6\rho\right]~.
\label{one_front_Q_more}
\end{equation}
This linear correction to the front speed is clearly apparent in the
plot of velocity versus $\rho$ in
Figure~\ref{front_dynamics_fig}a, obtained by direct numerical
solution of the PDEs.  Note also that beyond $\rho
\simeq 0.02$, even this next order asymptotic correction is
insufficient for quantitative accuracy.

{\it Interaction of two fronts.}  Reconsider the situation in an infinite
domain $-\infty \le x \le +\infty$ in which two interfaces are
symmetrically located at $x=\pm Q(t)$ between which $u=0$.  These
fronts are the edges of the localized state considered in the
heuristic derivation which obtained the dynamics
(\ref{Q_dynamics_heuristic}) -- this result is confirmed below using
the matched-asymptotic front speed formula (\ref{speed}).

The calculation proceeds as in the case of one front, but begins from
the outer activator pattern
\begin{equation}
u^0(x) = \cases{
0&$\mbox{}-\infty < x \le -Q(t)$~\cr
1&$\mbox{}-Q(t) \le x \le +Q(t)$~\cr
0&$\mbox{}+Q(t) \le x < +\infty$~\cr}
\label{u_outer2}
\end{equation}
that is symmetric about the origin.  Solving (\ref{outer_v}), the
outer inhibitor field can again be expressed in terms of exponential
functions
\begin{equation}
v^0(x) = \cases{
\sinh Q \;{\rm e}^{+x}& $\mbox{} -\infty < x \le -Q(t)$~\cr
1 - {\rm e}^{-Q}\;\cosh x&$\mbox{} -Q(t) \le x \le +Q(t)$~\cr
\sinh Q \;{\rm e}^{-x}&$\mbox{} +Q(t) \le x < +\infty$~.\cr}
\label{v_outer2}
\end{equation}
This determines the inhibitor contribution to the front speed
\begin{equation}
v^0(x=Q) = {1\over 2}\left(1-{\rm e}^{-2Q}\right)~,
\label{two_fr_v}
\end{equation}
where now, unlike in the case of a single front, $v^0(x=Q^j) \ne 1/2$ at the
interface.  Substitution into the asymptotic formula for the
front velocity (\ref{speed}) gives
\begin{equation}
Q_t \sim  - 6\sqrt{2D}~\left[\Delta F
- {\rho \over 2}{\rm e}^{-2Q}\right]
\label{two_fr_Q}
\end{equation}
where the minus sign must be taken since $u(x=Q^-)=1$.

The result (\ref{two_fr_Q}) not only recovers the heuristically-derived stripe
dynamics (\ref{Q_dynamics_heuristic}), but also determines the
constant of proportionality associated with the line tension $\gamma$
(\ref{gamma_define}) and the Rayleigh dissipation rate
(\ref{disspn_fnc_1d}).  Direct comparison of the two formulae
(\ref{Q_dynamics_heuristic}) and (\ref{two_fr_Q}) gives
\begin{equation}
	\gamma = {1\over 6}{\sqrt{D\over 2}}~,
\label{Gamma_exact}
\end{equation}
which is, as expected, ${\cal O}(\sqrt{D})$.  Moreover, since the line
tension coefficient (\ref{gamma_define}) represents the energetic
contribution proportional to perimeter, the leading order form of the
front profile $U^0$ can also be used in the integration
\begin{equation}
	\gamma = \sqrt{D}~\int_{-\infty}^{+\infty}\:d\eta~
		\left\{{1\over 2}~(U^0_\eta)^2 + F(U^0;1/2)\right\}~.
\label{line_tension_exact}
\end{equation}
Note that the additional contribution of $F(U;1/2)$ is a boundary
error associated with the pressure integral (\ref{energy_est}) and
resolves the discrepancy involving the line tension energy that was
alluded to below Eq. \ref{energy_est3}.

In addition to recovering the dynamics of stripe stabilization, the
single front speed formula can be used to understand the behavior of
two black stripes when $\Delta F > 0$.  If we were to start from a
configuration in which the stripes were not at their equilibrium
width, there would first be the ballistic collapse to form the black
stripes of width near $2Q^*$.  On a slower time scale, the two stripes
repel each other by their exponential tails of inhibition (on the
${\cal O}(1)$ scale) -- this is an illustration of the process of {\it
stripe repulsion} \cite{Ermentrout}.

\subsection{Numerical Comparison}

Tests of the results for the dynamics and equilibrium separation
of approaching fronts are shown in Fig. \ref{front_dynamics_fig}
for a situation like that in Fig. \ref{localized_state_fig}, but with
the values $D=0.0001$ and $\rho \le 0.02$, in the truly asymptotic
regime. The front separation $2Q$ is shown as a function
of time and is compared  with the predictions of Eq. (\ref{two_fr_Q}),
using both the bare prefactor $6\sqrt{2D}$ associated with the
leading-order profile, and the corrected prefactor $6\sqrt{2D}(1-6\rho)$
of Eq. \ref{one_front_Q_more}.  We see that the latter leads to highly
accurate results.  The equilibrium separation between the fronts is
independent of this prefactor, and as shown in Fig.
\ref{front_dynamics_fig}c is in accurate agreement with the results
of numerical solution of the full partial differential equations.

\section{Interface Motion in Two Dimensions}
\label{sec:TwoDim}

\subsection{Heuristic Derivation}

The heuristic derivation of the energetics and dynamics for contours
in two dimensions proceeds along the lines used for one-dimensional
patterns.  This requires the additional assumption, beyond that of
localized gradients, that the curvature of the boundary does not
significantly affect the interface profile.  We consider a single
black domain ${\cal B}$ of $u=1$ surrounded by a sea of white,
bounded by the contour ${\bf r}(s)$, where $s$
is arclength.  As in the calculation for
one dimension, the energy functional $\Delta {\cal E}$ consists of
three contributions.

The line tension contribution to the energy (\ref{slaved_energy}) must
now involve an integral around the boundary of the black
domain
\begin{equation}
\oint_{\partial{\cal B}}\!\! ds\gamma = \gamma L
\label{line_tension_est2d}
\end{equation}
where $L$ is the perimeter, and where the line tension $\gamma$
must be exactly that found earlier from the one-dimensional
asymptotics (Eq. \ref{Gamma_exact}).

By analogy with the result in Eq. \ref{energy_est}, the potential
integral contributes an energy proportional to the domain
area $A$,
\begin{equation}
\int_{\cal B}\!\! d{\bf x}\left[F(1;r)-F(0,r)\right] = \Delta F A~.
\label{energy_est2d}
\end{equation}

The main difference from the one-dimensional problem is
the treatment of the nonlocal term, which is proportional to the
double integral $\int\!\int\!{\cal G}$ over the domain.
Substituting ${\cal G}= \nabla^2{\cal G}+\delta({\bf x}-{\bf x}')$ we
may recast the bulk contributions as boundary terms by twice using the
Green's identity $\int_{A}\bbox{\nabla} \psi=\int_{\partial A}\hat{\bf
n}\psi$ in a manner analogous to that in Eqs. \ref{details1} and
\ref{energy_est2} as follows:
\begin{eqnarray}
\int_{{\cal B}}\!d{\bf x}\int_{{\cal B}}\! d{\bf x}' \! \ {\cal G} -A&=&
-\int_{{\cal B}}\! d{\bf x}\bbox{\nabla}_{{\bf x}}\cdot
\int_{{\cal B}}\!\! d{\bf x}'\bbox{\nabla}_{{\bf x}'}{\cal G}\nonumber \\
&=& -\int_{{\cal B}}\! d{\bf x}\bbox{\nabla}_{{\bf x}}\cdot
\oint_{\partial{\cal B}}\! ds'\hat{\bf n}(s'){\cal G}\nonumber \\
&=& -\oint_{\partial{\cal B}}\! ds\oint_{\partial{\cal B}}
\! ds'\hat{\bf n}(s)\cdot\hat{\bf n}(s'){\cal G}~.
\label{boundary_terms}
\end{eqnarray}
Note that $\hat{\bf n}(s)$ is the unit normal vector pointing {\it
out} of the black domain.  Finally, collecting together all of the
terms and noting that $\hat{\bf n}(s)\cdot\hat{\bf n}(s')=\hat{\bf
t}(s)\cdot\hat{\bf t}(s')$, where $\hat{\bf t}$ is the unit tangent
vector, we obtain
\begin{eqnarray}
\Delta{\cal E}[{\bf r}]&=&\gamma L + \Delta F A \nonumber \\
-{1\over 2}\rho\oint\! ds\oint\! ds'
\hat{\bf t}(s) \cdot \hat{\bf t}(s')
{\cal G}\left({\bf r}-{\bf r}'\right)~.
\label{energyfunc}
\end{eqnarray}
Unlike in one dimension, the nonlocality cannot be reduced to
a pointwise evaluation of the Green's function.

The form of the nonlocal coupling between tangent vectors is
reminiscent of the self-induction of current-carrying wires, the
direction of the current being specified by $\hat{\bf t}$.  Indeed,
the {\it sign} of the interaction is such that antiparallel tangent
vectors repel (like antiparallel current-carrying wires), while
parallel ones attract.
An alternative view of this connection with electromagnetic systems
is to consider the inhibitor relation (\ref{fast_inhibitor})
as a Poisson equation relating a potential ($v$) to a charge density
($u$), in which case the energy function ${\cal F}=\int uv$ in
Eq. \ref{energy_func} is the associated electrostatic energy.
Through the correspondence between
magnetization and current loops, and the similarity of electric and
magnetic dipolar phenomena, such a nonlocal interaction appears in the
description of pattern formation in a variety of other systems, as
detailed in Section XIII.  In such dipolar systems an important role
is played by molecular or other cutoffs in the self-induction
integrals.  The present formulation neither has such cutoffs nor needs
one for a well-defined energy, as the singularity in the Green's
function ${\cal G}$ in (\ref{energyfunc}) as $s-s'\to 0$ is
integrable, being only logarithmic in $\vert s-s'\vert$.

With the energy now formulated as a functional of the boundary contour
${\bf r}$, it only remains to calculate the Rayleigh dissipation
functional.  This is just the integral around the boundary of the
one-dimensional result
\begin{equation}
{\cal R}={\gamma \over 2D}\oint_{\partial{\cal B}}\! ds
\left(\hat{\bf n}\cdot {\bf r}_t\right)^2~.
\label{disspn_fnc_2d}
\end{equation}
To obtain the functional derivative of the energy (\ref{energyfunc}),
we combine the local contributions
\begin{eqnarray}
{1\over \sqrt{g}}{\delta {\cal R}\over \delta {\bf r}_t} & = & {\gamma \over D}
(\hat{\bf n}\cdot {\bf r}_t)\hat{\bf n}\nonumber \\
{1\over \sqrt{g}}{\delta L\over \delta {\bf r}} & = & \kappa\hat{\bf
n}\nonumber \\
{1\over \sqrt{g}}{\delta  A\over \delta {\bf r}} & = & \hat{\bf n}
\end{eqnarray}
where $g$ is the metric,
with the result for a self-induction integral \cite{Type_Ia,pra}
\begin{eqnarray}
{\delta \over \delta {\bf r}}{1\over 2}
\oint\! ds \oint\! ds'&&\hat{\bf t}\cdot \hat{\bf t}'{\cal G}(R) \nonumber \\
&&= \left\{\oint\!ds^\prime
\hat{\bf R}(s,s')\times \hat{\bf t}^\prime
{\cal G}'(R)\right\}\hat{\bf n}(s)~,
\label{self_induc_funcder}
\end{eqnarray}
where $\hat{\bf R}(s,s')=\left({\bf r}(s)-{\bf r}(s')\right)/ \vert
{\bf r}(s)-{\bf r}(s')\vert$, the cross product is a scalar in two
dimensions (${\bf a}\times {\bf b}= \epsilon_{ij}a_ib_j$), and the
prime on ${\cal G}$ indicates differentiation: ${\cal G}'\equiv d{\cal
G}/dR= -K_1(R)$.  The variational principle (\ref{gen_Lagrang}) determines
the contour evolution, expressed as the normal velocity
\begin{eqnarray}
\hat{\bf n}\cdot {\bf r}_t &=& - {D \over \gamma}
\biggl\{\gamma \kappa(s) + \Delta F \nonumber \\
&&\qquad - \rho \oint\! ds'
\hat{\bf R}(s,s') \times \hat{\bf t}^\prime{\cal G}'(R) \biggr\}~.
\label{normal_velocity}
\end{eqnarray}
Note that the nonlocal contribution in (\ref{normal_velocity}) is not
a singular integral, since the integrand for $s'\to s$ has the finite
limiting value
\begin{equation}
\lim_{s'\to s} \left\{ \hat{\bf R}(s,s')\times
K_1(\vert {\bf r}(s)-{\bf r}(s')\vert) \right\}
= - {1\over 2}\kappa(s)~.
\label{limiting_of_kernel}
\end{equation}

It is readily verified from Eq. \ref{normal_velocity} and Figure
\ref{Biot_Savart_fig} that the vector product in the normal velocity
embodies the repulsion between nearby antiparallel sections of the
contour.  As noted in earlier work \cite{prl}, this repulsion between
adjacent fronts requires only that ${\cal G}>0$ and ${\cal G}'<0$ (for
$\rho>0$), and not on the details of ${\cal G}$.

\subsection{Asymptotic Derivation of a Curvature Dynamics}

Now we rederive the contour dynamics (\ref{normal_velocity}) by
adapting the asymptotic argument presented earlier for one spatial
dimension.  The only major modification involves the reorientation of
the the inner coordinate ($\eta$) of the activator front to be in the
normal direction to the interface.

{\it Outer $O$(1) Scale:} In two dimensions, ${\bf x} = (x,y)$, the
leading-order outer representation satisfies the left-hand-side of the
equations
\begin{mathletters}
\label{out_2}
\begin{equation}
u(u-r)(u-1) = D\nabla^2 u-\rho(v-u)-u_t~,
\label{out_2u}
\end{equation}
\begin{equation}
\nabla^2v-(v-u)=\epsilon v_t~.
\label{out_2v}
\end{equation}
\end{mathletters}
The parameter scalings are chosen to obtain the analogous front
dynamics, but with the added dimensionality, the activator diffusion
scale ($\sqrt{D}$) is  incorporated into the significant limit
\begin{equation}
{Q_t \over \sqrt{D}} \sim \Delta F
\sim \rho \sim \sqrt{D} \ll 1~.
\label{signif_limit}
\end{equation}

As before, the underlying pattern is defined by patches where the
activator is in either of its bistable states
\begin{equation}
u({\bf x}) \sim u^0({\bf x}) = \cases{
1&$\mbox{black areas}$~\cr
0&$\mbox{white areas}$~.\cr}
\label{u2_out}
\end{equation}
Unlike in one-dimension where the transitions between black and white
are points, in two-dimensions the interface separating a black patch
from a white background is defined by a (slowly-moving) contour in
the $(x,y)$-plane
\begin{equation}
{\bf r}(\alpha;t) = (X(\alpha;t), Y(\alpha;t))
\label{curves}
\end{equation}
where $\alpha$ is defined as a counter-clockwise (but not necessarily
arclength) parameterization around the black patch.  There, of course,
may be more than one disjoint black patch in which case multiple
contours ${\bf r}^{\,j}$ must be evolved -- for simplicity of
presentation however, these indices are omitted.  The leading order
inhibition field, $v^0({\bf x})$, then satisfies the inhomogeneous
elliptic equation with a two-dimensional Laplacian
\begin{equation}
\nabla^2 v^0 - v^0 \sim - u^0 = - \cases{
1&$\mbox{black areas}$~\cr
0&$\mbox{white areas}$~.\cr}
\label{v2_out}
\end{equation}
Along the interface ${\bf r}$, continuity is imposed on $v^0$ and on
$({\bf \hat{n}} \cdot \nabla) v^0$, the normal derivative.  Again, with
suitable boundary conditions at the domain edges, a unique solution is
guaranteed.  For consistency with the separation of scales assumed in the
asymptotics, it is essential that the interface geometry contain
structure only on the inhibitor diffusion length -- this demands that
both the (global) interfacial separation distances and the (local)
radius of curvature be $O(1)$.

{\it Local Orthogonal Coordinates} The discontinuity of the activator on
the outer scale is naturally resolved on an inner scale which is a
stretched coordinate normal to the interface. At a point along the contour
${\bf r}(\alpha;t)$, the local (unnormalized) tangent $({\bf t})$ and
normal $({\bf n})$ directions are given by
\begin{eqnarray}
{\bf t}&=&{\bf r}_\alpha   \\
{\bf n}&=&{\bf J}{\bf r}_\alpha =
\left[ \begin{array}{cc}
0 & +1 \\ -1 & 0
\end{array} \right] {\bf r}_\alpha
\label{loc_vectors}
\end{eqnarray}
where the matrix ${\bf J}$ corresponds to a $90^\circ$ clockwise rotation.
With the counter-clockwise orientation of the parameterization $\alpha$,
the normal ${\bf n}$ points into regions of white. Using these as basis
vectors for a local coordinate system, introduce a new coordinate
$\bar{\eta}$ that extends orthogonally from the contour ${\bf r}$ in the
direction of the normal $({\bf n})$. This suggests the change of variable
(suppressing the time-dependence)
\begin{equation}
{\bf x}(\bar{\eta},\alpha) = {\bf r}(\alpha)
			+ \bar{\eta}{\bf J}{\bf r}_\alpha(\alpha)
\label{lin_change}
\end{equation}
so that $\bar{\eta}=0$ coincides with the contour ${\bf r}$.  In the
thin interface limit, the inner solution requires the validity of local
coordinates only for small $\bar{\eta}= O(\sqrt{D})$. It proves
calculationally more convenient to use instead an infinite (Lie) series as
the change of variable
\begin{eqnarray}
{\bf x}(\bar{\eta},\alpha) & = & {\bf r}
+ \bar{\eta}{\bf J}{\bf r}_\alpha
+{1\over 2}\bar{\eta}^2{\bf J}^2{\bf r}_{\alpha \alpha}+ \ldots \\
& \equiv & \exp \left\{\bar{\eta}\,{\bf J}
{\partial \over \partial\alpha}\right\} {\bf r}
\label{lie_change}
\end{eqnarray}
for which (\ref{lin_change}) is the linearization.  Straightforward
differentiation demonstrates that the complete series
(\ref{lie_change}) has the property that it is orthogonal even for
$\bar{\eta} \ne 0$ off the generating contour ${\bf r}$
\begin{equation}
{\bf x}_{\bar{\eta}} = {\bf J}{\bf x}_\alpha~.
\label{cauchy_r}
\end{equation}
These are just the Cauchy-Riemann conditions, thereby showing that the
change of variables between $(x,y)$ and $(\bar{\eta},\alpha)$ is locally
{\it conformal}. Although for small $\bar{\eta}$ only a few terms in this
series are necessary in the present analysis, it establishes to all orders
that the Laplacian nature of the diffusions are preserved
up to a Jacobian metric,
\begin{equation}
\nabla^2_{x,y} =
{1 \over \vert {\bf x}_\alpha(\bar{\eta},\alpha;t)\vert^2}
\nabla^2_{\bar{\eta},\alpha}~.
\label{laplacian_rescaled}
\end{equation}

{\it Inner $O(\sqrt{D})$ Scale:} The resolution of the activator
front, as well as the front speed determination now follows analogously to
the case of one dimension. The stretched inner coordinate $\eta$
\begin{equation}
\eta = {1 \over \sqrt{D}}{\bar{\eta}\over\vert{\bf r}_\alpha \vert}
\end{equation}
resolves the activator front across the discontinuity, while the
$\alpha$-coordinate labels position along the contour.
Using the expansion (\ref{lie_change}), we find that the metric factor in
(\ref{laplacian_rescaled}) is
\begin{equation}
\vert {\bf x}_\alpha(\bar{\eta},\alpha;t)\vert^2\simeq
\vert {\bf r}_{\alpha}\vert^2\left[1+2\sqrt{D}\eta \kappa + \cdots\right]~,
\end{equation}
where we have used the general expression for the curvature $\kappa$,
\begin{equation}
\kappa =
{\bf \hat{t}}\cdot{\bf \hat{n}}_s
= {{\bf r}_\alpha\cdot{\bf J}{\bf r}_{\alpha\alpha}
\over |{\bf r}_\alpha|^3}~,
\label{curv_def}
\end{equation}
defined such that it is positive for a
convex black domain.  Note that although the $|{\bf r}_\alpha|$-normalization
weakly breaks the orthogonality property, up to the first correction in
curvature
the Laplacian has the simple form
\begin{equation}
\nabla^2_{x,y} \sim {1\over D}
\left(1-2\sqrt{D}\eta\kappa\right)
{\partial^2 \over \partial \eta^2}+{\cal O}(1)~.
\label{lap_exp}
\end{equation}

At each point along the interface ${\bf r}(\alpha)$, the normal
component of the velocity is sufficient to describe the interface
motion.  This freedom essentially follows from an arbitrariness in the
contour parameterization $\alpha$, which permits arbitrary choices of
tangential velocity.  Here we may choose the
parameterization that yields, at fixed $\alpha$, an
interface moving normal to itself, with $Q_t(\alpha;t)$
the leading-order normal velocity,
\begin{equation}
Q_t(\alpha;t) = {\bf \hat{n}}\cdot{\bf r}_t~.
\end{equation}

At the level of the first correction, the inner equations for the front profile
are identical with the one-dimensional case (\ref{inner_equations}),
but with the addition of curvature term from the Laplacian expansion
(\ref{lap_exp})
\begin{mathletters}
\label{inner_eq2}
\begin{eqnarray}
{\cal S}[U] & \sim &	- {Q_t \over \sqrt{D}} U_{\eta}
+ 2\sqrt{D}\kappa\eta U_{\eta\eta}
- 6\Delta F U(U-1)\nonumber \\
&&+ \rho(V-U) + {\cal O}(D)~,
\label{inner_eq2u}
\end{eqnarray}
\begin{equation}
V_{\eta\eta} = {\cal O}(D)~,
\label{inner_eq2v}
\end{equation}
\end{mathletters}
where ${\cal S}$ is as defined above (\ref{L_define}). The leading-order
solution $U^0(\eta)$ is identical with the hyperbolic tangent solution
(\ref{hyp_tan}). Application of the same solvability inner product
(\ref{solve}) yields the front speed relation
\begin{eqnarray}
Q_t&\sim& -6\sqrt{2D}~\Biggl\{{1\over 6}\sqrt{D \over 2} \kappa
+ \Delta F \nonumber \\
&&\qquad\qquad+\rho\,\left[v^0(x=Q)-{1 \over 2}\right]\Biggr\}~,
\label{speed_2}
\end{eqnarray}
which clearly recovers the curvature and potential effects of the
energetically-determined front velocity (\ref{normal_velocity}).  In
order for the two expressions to be in complete agreement, the
nonlocal contribution of the contour integral in
(\ref{normal_velocity}) must then be equivalent to the contribution
from the inhibition term $v^0(x=Q)$ in (\ref{speed_2}).

The nonlocal character of the inhibitor field on the pattern is
expressed mathematically by the Green's function solution of the
outer inhibitor equation (\ref{v2_out})
\begin{equation}
v({\bf x}) = -{1 \over 2 \pi}\,\int_{\cal B}\! d{\bf x}'
K_0(|{\bf x}-{\bf x'}|)~,
\label{green2d}
\end{equation}
where the integration is done only over black areas. Using the PDE for
$K_0$ and Green's theorem, the above solution can be rewritten in terms of
a line integral over the interfacial boundary ${\bf r}$. This results in
the arclength integral
\begin{equation}
v({\bf r}) = {1 \over 2} + {1 \over 2 \pi}\oint\! ds^\prime\,
{\bf \hat{n}} \cdot{{{\bf r}-{\bf r'} \over \vert{\bf r}-{\bf r'}\vert}}
K_1(|{\bf r}-{\bf r'}|)~,
\label{green_conta}
\end{equation}
where $K_1$ is introduced via the Bessel identity
$K_1=-K_0^\prime$. It should be noted that a Plemelj-type argument
\cite{Carrier} is required to evaluate this integral since $v^0$ must
be evaluated for a point on the boundary ${\bf r}$.  Here however, the
usual principal value integral is unnecessary because the singularity
in the integrand is removable (see Eq. \ref{limiting_of_kernel}).

Substitution of the contour integral (\ref{green_conta}) into
the asymptotic front speed (\ref{speed_2}) confirms the energetic
result (\ref{energyfunc}).  In terms of $v^0(x=Q)$, the
global energy of a black pattern may be written as
\begin{equation}
{\cal E} = \gamma L +
{\rho \over 2}\int_{\cal B}\! d{\bf x}'
\left[v^0({\bf x})-1+{2\Delta F\over \rho}\right]
\end{equation}
a result that is not immediately obvious from the asymptotic
derivation.
Written in this way, with the area $A$ written as
an integral, we see immediately from the fact that $v^0-1 \le 0$
that the integrand loses positive definiteness for
$\rho > 2\Delta F$, which is precisely the criterion for
the stability of stripes (cf. Eq. \ref{L_equil}).

As formulated here, the front velocity formulas
(\ref{speed_2}) and (\ref{energyfunc}) both involve the
evaluation of the outer inhibitor field, either by boundary
integration (\ref{boundary_terms}), or direct solution of the bulk PDE
(\ref{v2_out}).  The boundary approach offers a compact, intrinsic
description of complex contours; however, the bulk PDE approach can be
calculationally advantageous in simple geometries where natural
eigenfunctions can be constructed.  An example of the latter is the
analysis of a disc-shaped domain.

{\it Localized Disk Solution:} For a black circular spot of radius $R$
in an infinite white domain, the outer inhibitor $v^0$ has a
radially-symmetric solution of (\ref{v2_out}) in terms of modified
Bessel functions
\begin{equation}
v^0(r)=\cases{
1 - R\,K_1(R)\,I_0(r)&$r \le R$~\cr
R\,I_1(R)\,K_0(r)&$R \le r$~\cr}~,
\label{v_inout}
\end{equation}
where continuity of $v^0$ and $v^0_r$ is satisfied at $r=R$.  Direct
substitution of $v^0(R)$ into (\ref{speed_2}) gives a nonlinear ODE
for the purely radial evolution of the contour
\begin{eqnarray}
R_t &\sim& -6\sqrt{2D}\Biggl\{{1\over 6}\sqrt{D \over 2}{1\over R}+\Delta F
\nonumber \\
&&\qquad\qquad+ \rho\left[RI_1(R)K_0(R)-{1 \over 2}\right]\Biggr\}~.
\label{radius_evolution}
\end{eqnarray}

The dynamics (\ref{radius_evolution}) shows that the condition
for the existence and radial stability of equilibrium disk solutions
depends only on two ${\cal O}(1)$ parameters: the potential difference
and the inhibitor coupling, both scaled on the activator diffusion length.
Equilibrium radii $R^*$ then satisfy the transcendental
condition
\begin{equation}
{\sqrt{D}\over 6\sqrt{2}R^*}+\Delta F
+\rho
\left[R^*I_1(R^*)K_0(R^*) - {1 \over 2}\right] = 0
\label{disc_radius}
\end{equation}
which, upon numerical root tracking, identifies four distinct
equilibrium scenarios.  These are illustrated in Figure~\ref{circle_fig}:

\indent{\bf (a)} two equilibria, only larger radius stable;

\indent{\bf (b)} one equilibrium, unstable;

\indent{\bf (c)} no equilibria, radial contraction only;

\indent{\bf(d)} three equilibria, intermediate disk stable.

\noindent Apart from the anomalous region {\bf (d)}, note that black
localized disk patterns can only be stabilized when $\Delta F>0$ and
white is energetically-preferred.  This behavior is the analog of the
one-dimensional stripe stabilization and illustrates the process of
{\it domain localization}.  Although stable disks can be sustained in
the small region {\bf (d)} for $\Delta F<0$, since this
region is so small in parameter space and the attracting basin for the
stable radius relatively narrow, we believe these states to be
unimportant, at least, in the context of this particular formulation
of the FitzHugh-Nagumo system.

It is also important to note that in the absence of inhibitor
coupling, no stable disks are possible.  The case when $\Delta F =0$
(and $\rho=0$) corresponds to the well-known shrinkage by curvature
that leads to a singular collapse in finite time
\begin{equation}
R(t) \sim \sqrt{2D (t_0-t)}~.
\label{shrink}
\end{equation}

Note that the PDE evolutions shown in Figs.~{\ref{intro_sim_fig1}} and
{\ref{intro_sim_fig2}} both have parameter values that lie within
region {\bf (a)}.  The dynamical evolution in
(\ref{disc_radius}) is restricted to radially-symmetric geometries;
the possibility for azimuthal instabilities, which ultimately generate
the labyrinthine patterns, is addressed in the following section.

Just as we obtained the energy of a domain in one dimension and
deduced the existence of an energy barrier to its creation, we may
compute the energy of a circular domain of radius R on the basis of
Eq. \ref{energyfunc}.  The computation of the nonlocal contribution
may be done two ways, via the direct calculation of the self-induction
integral, or by finding the inhibitor field $v(r)$ associated with a
circular activator pulse.  Collecting all of the contributions, we
obtain the energy of a circle of radius $R$,
\begin{equation}
\Delta{\cal E}(R) = 2\pi R\gamma + \pi R^2 \Delta F
- \pi\rho R^2K_1(R)I_1(R)~,
\label{circle_energy}
\end{equation}
Using the simplest estimate of $\gamma$ from Eq. \ref{Gamma_exact},
Figure \ref{circle_energy_fig}
shows for several different values of $\rho$ the function
$\Delta{\cal E}(R)$ in Eq. \ref{circle_energy}, illustrating the presence of a
local
minimum at a finite value of $R$ for sufficiently large $\rho$.

\section{Instabilities of Chemical Fronts}
\label{sec:Instab}

\subsection{Mechanism of the transverse front instability}

The last of the three important aspects of the reaction-diffusion dynamics
outlined in the introduction concerns the instabilities of
chemical fronts.
At the level of linear stability analysis, we will see that the mechanism
of this instability is rather similar to that of the Mullins-Sekerka
instability of a liquid-solid interface \cite{Mullins_Sekerka}.
In Figure \ref{front_instability_fig} we
show schematically the level curves of activator and inhibitor concentration
near a modulated interface of the inhibitor.
The excess of inhibitor in the concave regions of the front has
the tendency to push those concavities further back.  Conversely, the
portions of the front that are convex outwards are locally depleted in
inhibitor
relative to the flat interface, and will expand further outward.  Both of these
effects increase the length of the interface and its curvature, and will
thus be resisted by line tension.  The two effects will balance at
a characteristic length scale to produce an instability.

The instability may be understood in its simplest form at the level of
the energetics of circular domains;  consider again the expression
in Eq.(\ref{circle_energy}) and examine the local limit $1/R\ll 1$
In this limit, we use the asymptotics
$K_1(z)I_1(z)\simeq(1/2z)(1-3/8z^2 + \cdots)$ to simplify the energy
\begin{equation}
\Delta{\cal E}=2\pi R\left(\gamma -{1\over 4}\rho\right)+\pi R^2\Pi
+{3\pi\rho\over 16 R} + \cdots~.
\label{energy_limit}
\end{equation}
We see two consequences of the inhibitor coupling.  First there is an
effective line tension $\gamma-\rho/4$ which may be driven negative
for sufficiently large activator-inhibitor coupling.  Clearly a negative
line tension will lead to proliferation of the interface.
A negative line tension is a common
feature of fingering instabilities arising from the competition
between Young-Laplace and Biot-Savart interactions \cite{science,pra,pre,jpc}.
Second, at higher-order
in $1/R$ there appears a stabilizing contribution which below we show is
like that for the bending energy of
an elastic rod.  If the effective tension is negative, then this last
contribution will stabilize the interface against small-scale disturbances
and produce a finite-wavelength instability.

\subsection{Approximate Local Contour Dynamics}

The heuristic notion of an effective line tension may be made more systematic
by recasting the nonlocal contour dynamics in an approximate local
(but still intrinsic) form through
an expansion in powers of the presumed small curvature $\kappa$,
and its arclength derivatives.  The method
of this expansion is similar in spirit to that used recently
\cite{Duplantier} to study screened electrostatic potentials near
surfaces of arbitrary geometry.  Indeed, the inhibitor field obeys
the same modified Helmholtz operator as appears in the Debye-H\"uckel
theory of colloidal interactions \cite{Verwey}, which leads to the
screened Coulomb interaction between elementary charges.

The curvature expansion may be constructed directly from the equation
of motion, or first at the level of the energy functional and then
carried through the variational principle.  Adopting the latter method,
we first expand the scalar product of tangent vectors in the self-induction
integral (\ref{energyfunc}) as a power series in $\Delta =s'-s$,
\begin{equation}
\hat{\bf t}(s)\cdot \hat{\bf t}(s')\simeq 1-{1\over 2}\Delta^2\kappa^2
+\cdots~,
\label{that_expand}
\end{equation}
and the distance function
\begin{equation}
\vert {\bf r}(s')-{\bf r}(s)\vert\simeq \vert\Delta\vert
-{1\over 24}\Delta^2
\vert \Delta\vert \kappa^2 + \cdots~.
\label{dist_expand}
\end{equation}
Substituting into the self-induction integrals and extending the limits of
integration over $s'$ to $\pm \infty$ (with exponential accuracy), we obtain
\begin{equation}
\oint\! ds \oint \! ds' \hat{\bf t}\cdot \hat{\bf t}'
K_0\simeq
\pi\oint\! ds \left(1-{3\over 8}\kappa^2(s) +
\cdots\right)~,
\label{integs_expand}
\end{equation}
and the approximate local energy functional
\begin{equation}
\Delta{\cal E}[{\bf r}]\simeq \Pi A +  {\bar \gamma} L
+{3\over 32}\rho\oint\! ds \kappa^2~,
\label{local_energy_contour}
\end{equation}
where the effective line tension is as in Eq. \ref{energy_limit}:
${\bar \gamma}=\gamma-(1/4)\rho$.
The term proportional to $\kappa^2$ in Eq. \ref{local_energy_contour}
is the bending energy of an elastic line \cite{Landau_elasticity}, and its
coefficient is positive if the activator-inhibitor coupling constant $\rho$ is
positive.  Under that condition, it prevents the interface from bending on
arbitrarily fine scales.

Using the approximate local energy functional (\ref{local_energy_contour}),
the normal velocity of the interface is
\begin{equation}
{\gamma\over D}\hat{\bf n}\cdot {\bf r}_t\simeq -\Delta F-{\bar \gamma} \kappa
+{3\over 16}\rho\left(\kappa_{ss}+{1\over 2}\kappa^3\right)~.
\label{geometric_model}
\end{equation}
Apart from the pressure term, this is the planar version of the
``curve-straightening equation" \cite{Griffiths} that is
equivalent to the Rouse dynamics of the worm-like model of elastic
polymers \cite{folding}.
It has the form of a Landau expansion in powers of the curvature and its
derivatives, and in that sense is similar to so-called ``geometric" models of
crystal growth \cite{Brower}.  Unlike those models, however, the coefficients
of
the various terms in Eq. \ref{geometric_model} are not all independent.  Thus,
for instance, the terms $\kappa_{ss}$ and $\kappa^3$ must have coefficients
with ratio $2$ by virtue of the variational principle applied to the
energy functional (\ref{local_energy_contour}).

\subsection{Linear stability for fronts and stripes}

The most fundamental instability of chemical fronts in the present model is
that of an infinite straight interface bounding the states $u=0$ and $u=1$.
To compute the spectrum of growth rates of perturbations to this front,
we parametrize the interface as
\begin{equation}
{\bf r}(s)=x\hat{\bf e}_x + \zeta(x,t)\hat{\bf e}_y~,
\label{interface_param}
\end{equation}
If we let $\zeta(x,t)=\zeta_k(t) \cos(kx)$, the linearization of the
Biot-Savart integral is
\begin{eqnarray}
\oint\!ds'{\hat{\bf R}}
\times\hat{\bf t} K_1&\simeq&
\zeta_k\cos(kx)\times \nonumber \\
&& \int_0^{\infty}\!\! {dy \over y}
(k y\sin(2k y)-2)K_1(y)~,
\label{Biot_linearized}
\end{eqnarray}
Use of the identity $(1/y)K_1(y)= -K_0(y)-K_1'(y)$ and an integration by parts
transforms (\ref{Biot_linearized}) into a standard integral.  Combining these
results with the linearized curvature, the
growth rate $\sigma(k)=(\partial_t\zeta_k)/\zeta_k$ for the single front
instability is
\begin{equation}
\sigma_{\rm f}(k)= 6\sqrt{2D}\left(-\gamma k^2
+ {\rho\over 2} \left[1-{1\over f(k)}\right]\right)~.
\label{buckling_sigma}
\end{equation}
where
\begin{equation}
f(k)=\sqrt{1+k^2}~.
\label{fofk_define}
\end{equation}
For wavelengths that are small relative to the ${\cal O}(1)$ inhibitor
screening length, we obtain
\begin{equation}
\sigma_{\rm f}(k)\simeq 6\sqrt{2D}\left(-\left[\gamma-{1\over 4}\rho\right]k^2
-{3\over 16}\rho k^4 + \cdots\right)~,
\label{buckling_expand}
\end{equation}
showing, as discussed in the previous section (see in particular Eq.
\ref{geometric_model}), that the instability arises from a
negative effective line tension, stabilized by an elastic-type term in $k^4$.
This result also shows that at the onset of the instability the critical
wavelength is infinite.  As in the analysis of
steady states for disk solutions, Eq. \ref{disc_radius}, we shall find it
convenient to
work in a rescaled parameter space with coordinates $\tilde\rho$ and
$\tilde r$ as in (\ref{rescaled_rho_r}).
Setting $\sigma(k)=0$ and $d\sigma(k)/dk^2=0$ we obtain
the critical coupling constant for the instability,
\begin{equation}
\tilde \rho_f= {4\gamma\over \sqrt{D}} = {\sqrt{2} \over 3}~.
\label{Omega_crit_buckling}
\end{equation}
Note that this is independent of the parameter $r$, and hence also of
the energy difference $\Delta F$, and so is equally valid as a stability
criterion for a uniformly moving straight front.
The critical value of $\rho$ could also be obtained directly from the vanishing
of the
effective line tension in (\ref{buckling_expand}).  For $\rho > \rho_f$,
the most unstable wavevector is
\begin{equation}
k^*=\left[\left({\rho\over \rho_f}\right)^{2/3}-1\right]^{1/2}~.
\label{unstable_k_buckling}
\end{equation}
Figure \ref{stripe_stability_fig} shows in the rescaled parameter space
the location of the instability of a single front, as well as the
boundary of stable stripes, defined by
\begin{equation}
\tilde \rho={1\over 3} \tilde r~.
\label{stripe_existence}
\end{equation}

Beyond the behavior of a single chemical front, it is natural to consider
the stability of stripes.  Perturbations to the shape of a stripe may
be decomposed into those with odd and even parity under reflection through
the stripe midline, termed {\it sinuous} (S) and {\it varicose} (V),
respectively.
In these two situations, the interfaces bounding a stripe of width $2Q^*$ are
parametrized as follows:
\begin{mathletters}
\label{stripe_perts}
\begin{equation}
{\bf r}_{\pm}(x)=x\hat{\bf e}_x
+ \left(\pm Q^*+\zeta(x,t)\right)\hat{\bf e}_y \ \ \ \ \
({\rm S})~,
\label{interface_param_stripe_serp}
\end{equation}
\begin{equation}
{\bf r}_{\pm}(x)=x\hat{\bf e}_x
+ \left(\pm Q^*\pm\zeta(x,t)\right)\hat{\bf e}_y \ \ \ \ \
({\rm V})~.
\label{interface_param_stripe_peri}
\end{equation}
\end{mathletters}
A straightforward calculation yields the growth rates of these two modes:
\begin{mathletters}
\label{stripe_growth}
\begin{eqnarray}
\sigma_{\rm S}(k)&=& \sigma_{\rm f}(k)-3\sqrt{2D}
\rho{\rm e}^{-2Q^*}\nonumber \\
&&\qquad\qquad\times \left(1-{1\over f(k)}{\rm e}^{-2Q^*(f(k)-1)}\right)~,
\label{stripe_growth_S}
\end{eqnarray}
\begin{eqnarray}
\sigma_{\rm V}(k)&=& \sigma_{\rm f}(k)-3\sqrt{2D}\rho
{\rm e}^{-2Q^*}\nonumber \\
&&\qquad\qquad\times \left(1+{1\over f(k)}{\rm e}^{-2Q^*(f(k)-1)}\right)~.
\label{stripe_growth_P}
\end{eqnarray}
\end{mathletters}
In each case, the growth rates are less than that of the single front.
The sinuous mode grows faster than the varicose, whose damping
at $k=0$ reflects the local stability of the preferred stripe
width. This is consistent with what is known in other contexts, for
instance in the buckling instability of magnetic stripes
\cite{Cebers}.  Figure \ref{front_stability_fig} shows these growth
rates for typical values of the parameters.  The greater stability of
a stripe relative to a single front may also be revealed by the onset
of its sinuous instability at higher values of $\rho$.  Expanding
the growth rate (\ref{stripe_growth_S}) to order $k^4$ as in
Eq. (\ref{buckling_expand}), we obtain marginal stability conditions
by setting the coefficient of $k^2$ to zero, yielding the
transcendental equation
\begin{equation}
{\sqrt{2}\over 3}=\tilde\rho\left\{1-{\tilde r\over 3\tilde \rho}
\left[1+\ln\left({3\tilde \rho\over \tilde r}\right)\right]\right\}~.
\label{stripe_serp_inst}
\end{equation}
A numerical solution to this is shown in Fig. \ref{stripe_stability_fig}, where
we see
that this stability boundary properly merges with that of a single front as
$r\to 1/2^+$,
and the equilibrium stripe width $Q^*$ diverges.
Note that Eq. \ref{stripe_serp_inst} is precisely the condition that the
structural energy per unit length of the stripe (\ref{delta_E_nifty}) vanish.
That these two conditions are related is seen by the fact that the coefficient
of $k^2$ in the stability analysis is an effective line tension or
energy per unit length. Beyond this stability boundary, the proliferation
of stripes (straight or buckled) is energetically preferred.

\subsection{Azimuthal Instability of Disks}
The evolution equation (\ref{radius_evolution}) that establishes the
size of equilibrium disk solutions (\ref{disc_radius}) also determines
their nonlinear stability to radial perturbations.  For azimuthal
disturbances, the growth exponents are obtained by considering
a disk whose radius develops a small sinusoidal
variation
\begin{equation}
r = R^* + \zeta(t)\sin n\theta~.
\label{disc_pert}
\end{equation}
The linearized curvature of the interface as calculated using
(\ref{curv_def}) is
\begin{equation}
\kappa \sim {1\over R^*} + (n^2-1){ \zeta(t)\over {R^*}^2}\sin n\theta~,
\label{disc_pert_curv}
\end{equation}
where it is noted that the $n=\pm 1$ perturbations are equivalent to
simple translations of the disc.  In this near-circular geometry, it is
advantageous to use the bulk approach to solve for the inhibitor field
since the leading-order correction to the unperturbed inhibitor field
involves only a single eigenmode.  To ${\cal O}(\zeta)$, the inhibitor field
$v^0(r)$ is
\begin{equation}
v^0_n(r,\theta) \sim v^0(r) + \cases{
\zeta R^*K_n^*I_n(r)\sin n\theta&$r\le R$~\cr
\zeta R^*I_n^*K_n(r)\sin n\theta&$R\le r$~\cr}~,
\label{pert_inh1}
\end{equation}
where $v^0(r)$ is given in (\ref{v_inout}), $K_n^*\equiv K_n(R^*)$ and
$I_n^*=I_n(R^*)$,
and where the continuity conditions must now be imposed on the perturbed
interface.  This gives the corrected value of the inhibitor on the
interface
\begin{equation}
v^0_n({\bf r}) = v^0_r(R^*)
-\zeta R^*\left[K_1^*I_1^*-I_n^*K_n^*\right]\sin n\theta~,
\label{pert_inh2}
\end{equation}
where the correction also vanishes for the $n=\pm 1$ modes.

Collecting the ${\cal O}(\zeta)$ corrections from the front speed
dynamics (\ref{speed_2}) yields the linearized growth exponent for
the $n^{\rm th}$-mode perturbation
\begin{eqnarray}
\sigma_n= 6\sqrt{2D}\left\{\gamma{1-n^2 \over {R^*}^2}
+\rho R^*\left[K_1^*I_1^*-I_n^*K_n^*\right]\right\}~.
\label{azi_exp}
\end{eqnarray}
The neutral curves for the lower mode instabilities acting on the disc
solutions of region {\bf (a)} are shown in Figure~\ref{circle_fig}.
Coincidentally, note that the convergence point for these curves is
identical to the critical $\rho_c$ (\ref{Omega_crit_buckling}) which
is where $R^*$ is naturally large.  The fact that the instabilities of
disks and straight fronts derives from a common mechanism is
demonstrated in the limit of large disks yet finite wavenumber
$k = n/R^*$, yielding
\begin{equation}
\lim_{n,R^* \to \infty} \sigma_{(k=n/R^*)} = \sigma_f(k)~.
\label{azi_lim}
\end{equation}

\section{Numerical Studies}
\label{sec:Numerics}

\subsection{Simulation of PDEs}

In this section we present numerical studies of the
reaction-diffusion dynamics (\ref{uvequations2}) to verify the stability
results
discussed in the previous section, within the context of the
parameter space of Fig. \ref{stripe_stability_fig}.
In the subsequent section we investigate the contour dynamics for comparison.
These numerical studies were performed using a pseudospectral algorithm
in a two-dimensional periodic domain.  The algorithm is outlined in Appendix A.
In showing the time evolution it is convenient to adopt a rescaled
time $\tau\equiv \sqrt{2D}(r-1/2) t$

The first phenomenon illustrated is the two-dimen-\break sional version of the
approach in one dimension of two chemical fronts illustrated in Figure
\ref{localized_state_fig}.
This is shown in Fig. \ref{2dfronts_fig} starting from an initial condition
in which the two fronts are far apart and given a modulation by a random
collection of Fourier modes.  To indicate the symmetry of the underlying
PDEs given in Eq. \ref{symmetries}, the parameters chosen ($D=0.01$,
$\rho=0.2$,
$r=0.4$) are such that the state $u=1$ is the more stable and
invades the regions with $u=0$.  The straight front is linearly stable,
so in addition to the net motion of the fronts toward each other,
these initial modulations relax.
As in the one-dimensional studies, for these parameter values the fronts
do not cross, coming to rest at a finite distance set by the detuning
$r-1/2$ and the inhibitor coupling $\rho$.
Similar behavior has been seen in
the iodide-ferrocyanide-sulfate reaction \cite{Lee}, as
well as in simulations of the Grey-Scott model \cite{Lee_long}.

The growth of a labyrinthine structure from a compact initial condition
was illustrated in Fig. \ref{intro_sim_fig1}.  The parameter values for this
computation, $\tilde r=0.2$, $\tilde \rho=1.5$, are in the region of rescaled
parameter space in which the simple disk solution is unstable to many
azimuthal modes.  The black regions in the figure are those within which
$u\ge 1/2$.
We see in the early stages of the
evolution the growth of fingers of a
well-defined width -- this is the behavior discussed by Ohta et al.
\cite{Ohta}.
Several of the fingers undergo tip-splitting.
Their mutual repulsion leads eventually to
a space-filling labyrinth which apparently converges to a steady state.
This convergence is clearly a consequence of the front self-avoidance in the
periodic computational domain.
The interactions between the fronts have been such as to
create a rather uniform width to the fingers of $u=1$, as well as to the
intervening regions with $u=0$.
Besides their similarity to the experimental patterns of Lee, {\it et al.}
\cite{Lee,Lee_long}, the phenomenology of this pattern formation has a very
strong resemblance to that seen in magnetic fluids \cite{Rosensweig},
superconductors \cite{Huebener}, as well as thin garnet films
\cite{Seul_films}.

By changing the coefficients to $\tilde r=1$, $\tilde \rho=1$, we enter
the regime in which the localized disk is stable, and
a branched structure may relax to it without fissioning.  This
is shown in Fig. \ref{intro_sim_fig2}, in which the starting field
configuration
is panel (c) in Fig. \ref{intro_sim_fig1}.
Again, this shape relaxation is like that seen in the
iodide-ferrocyanide-sulfite reaction, as well as that observed in
magnetic fluids in Hele-Shaw flow when the applied magnetic field is removed
and surface tension returns a fingered structure to the circular ground
state \cite{science}.

The front interactions responsible for labyrinthine fingering
instabilities of a single domain naturally appear in the interactions
of multiple domains, and can lead to patterns composed of disconnected
but highly interdigitated regions.
Figure \ref{multiple_domains_fig} shows contour plots of the evolution
starting from an initial condition with two small, nearly circular
domains.  Despite the complex fingering instabilities, the domains
retain their integrity and do not merge.  This multiple-domain problem
is rather similar to that observed in the intermediate state of Type-I
superconductors (see also Section \ref{sec:Nonlocal}).

An important aspect of the evolution of compact domains concerns the
possibility of domain fission and fusion.  As is quite common with interfacial
treatments of pattern formation, the contour dynamics is
not asymptotically valid when regions of the interface approach
each other near the reconnection point point. The full
PDEs for the activator-inhibitor pair are however well-defined during
these events, and both fission and fusion are possible if the pressure
driving the interfaces together is large enough, or conversely the
inhibitor-induced repulsion is small enough.
This has also been observed in experiments \cite{Lee_long}.

\subsection{Simulation of contour dynamics}

Next we turn to numerical studies of the nonlocal contour dynamics.
The numerical method for this, outlined in Appendix A, has been
employed elsewhere \cite{jpc} for analogous problems in nonlocal interface
motion.  It is based on a pseudospectral treatment of the tangent-angle
representation of the dynamics.

The most elementary instability of a localized state is the elliptical
one, illustrated in Fig. \ref{circle_to_stripe_fig}.  Here, the
initial condition is a solution of the equilibrium condition
(\ref{disc_radius}) for a localized disc, perturbed with a small
amplitude of the $n=2$ mode.  After a short transient, the domain
enters an era of linear elongation as illustrated in the lower panel
of the figure.  This linear growth is a reflection of the negative
structural energy density (\ref{delta_E_nifty}) for the parameters
$\tilde r=0.212$, $\tilde \rho=0.796$.

Increasing the value of the inhibitor coupling to $\rho=0.90$ renders the
localized state unstable to the mode $n=3$ (as well as to the elliptical
mode).  The bulbous tips that form during the elliptical instability have
a nonzero projection onto the $n=3$ mode, leading to tip-splitting events.
This is shown in Fig. \ref{tip_splitting_fig}.  In this regime of
parameter space each time a new tip is
formed it is susceptible to tip-splitting, leading to a cascading process
and a proliferation of three-fold coordinates vertices.

For $\rho$ larger still, we may investigate one of the common features
of labyrinthine structures seen in dipolar systems as well as in
chemical reactions: the appearance of three-fold nodes to the
exclusion of all others of higher coordination.  For certain kinds of
optimization problems involving the minimization of interface length
under the constraint of fixed endpoints it is known that three-fold
nodes are the only stable vertices \cite{Steiner}, but no such result
is known in the present context. The numerical simulations strongly
suggest, however, that higher-order vertices are dynamically unstable.
Figure \ref{vertex_fig} shows the evolution of an initial condition
consisting of a localized disk modulated by four-fold perturbations of
several per cent and a two-fold distortion one fiftieth as large.  A
cross-like vertex forms quickly, but is unstable to the elliptical
perturbation, splitting into two three-fold vertices.

Having seen in isolation the elementary processes underlying labyrinth
formation, stripe formation and proliferation, tip-splitting, and
vertex reduction, we show in Fig. \ref{circle_to_labyrinth_fig} the
appearance of a labyrinthine pattern from a compact initial condition.
The characteristic feature of interface repulsion and the appearance
of well-defined finger widths is readily apparent.  Unlike the
simulations of the PDEs, the contour representation does not naturally
build in periodic boundary conditions, so the long-time evolution of
the two will differ significantly.  Changing the coefficients so that
the localized disk is both radially and azimuthally stable, we see in
Fig. \ref{labyrinth_relax_fig} the relaxation to a compact state of a
branched initial condition taken from near the the end of the
evolution in Fig. \ref{circle_to_labyrinth_fig}.

\section{Beyond the Slaving Limit}
\label{sec:Memory}

As we have discussed so far, the limit $\epsilon\to 0$ renders the dynamics an
overdamped gradient flow, associated with a Lagrangian variational principle
in which the kinetic energy is neglected and the dissipation function is
local.  As the recent work of Hagberg and Meron \cite{Meron,Hagberg}
has emphasized, the nature of the chemical fronts between the two
metastable values of the activator may be quite different in the
two extremes of slow- and fast-inhibition.  From Eq. \ref{speed_2},
the timescale of the front dynamics is $u_t=O(D)$, so that in actuality,
the fast-inhibitor limit only requires that $\epsilon\,D \ll 1$ -- so that
the value of $\epsilon$ can be quite large.  Extending this contour dynamic
approach beyond this fast-inhibitor limit represents a
significant challenge.  Nevertheless, an intuition for
how oscillatory behavior might arise with finite $\epsilon$ is suggested by
arguments using contour energetics.

For $\epsilon \ne 0$ we may solve the
inhibitor dynamics (\ref{uvb2model}) for $v$ in terms of $u$,
\begin{equation}
v({\bf x},t)=\int^t\!\! dt' \int\! d{\bf x}' G({\bf x}-{\bf x}',t-t')u({\bf
x}',t')~,
\label{v_solve}
\end{equation}
where for $t>0$ the Green's function is
\begin{equation}
G({\bf x},t)={1\over 4\pi t}\exp\left\{-{t\over\epsilon}
-{\epsilon{\bf x}^2\over4 t}\right\}~.
\label{full_Green}
\end{equation}
Clearly, $\epsilon$ is the natural time scale for the decay of $G$, and if we
take the liberty of expanding the integral for
slowly-varying $u$ we obtain
$u({\bf x}',t')\simeq u({\bf x}',t)+(t'-t)u_t({\bf x}',t) + (1/2)(t'-t)^2
u_{tt}({\bf x}',t) + \cdots$,
substitute into (\ref{full_Green}) and perform the time integrations.  Each
power of
$t'-t$ in the expansion will contribute a power of $\epsilon$.  Up to order
$\epsilon^2$ we obtain
\begin{eqnarray}
v({\bf x},t)&\simeq& \int\! d{\bf x}'\biggl\{{\cal G}_0({\bf x}-{\bf x}')
u({\bf x}',t)
+ \epsilon {\cal G}_1({\bf x}-{\bf x}') u_t({\bf x}',t)\nonumber \\
&&\qquad ~~~ + \epsilon^2 {\cal G}_2({\bf x}-{\bf x}')
u_{tt}({\bf x}',t)\biggr\}~,
\label{v_time_expand}
\end{eqnarray}
where ${\cal G}_0({\bf x})=(1/2\pi)K_0\left(x\right)$ and
the remaining functions $G_i$ are
\begin{equation}
{\cal G}_1({\bf x})= -{1\over 4\pi}{x}K_1\left({x}\right)~, \ \
{\cal G}_2({\bf x})={1\over 16\pi} x^2 K_2\left({x}\right)~.
\label{Gi_defs}
\end{equation}

Note the very important feature that the order $\epsilon$ kernel ${\cal G}_1$
is
negative, while the second-order kernel is positive.
Up to quadratic order in $\epsilon$ the nonlocal activator dynamics may be
written in the following form
\begin{eqnarray}
\epsilon^2\rho\int\! d{\bf x}' {\cal G}_2&&({\bf x}-{\bf x}') u_{tt}({\bf
x}',t)
+{\delta {\cal E}\over \delta u} = \nonumber \\
&&-u_t-\epsilon\rho\int\! d{\bf x}' {\cal G}_1({\bf x}-{\bf x}') u_t({\bf
x}',t)~.
\label{u_dynamics_eps2}
\end{eqnarray}
This conforms to the variational principle in Eq. (\ref{gen_disspn_fnc_1d}),
having the form
\begin{equation}
\partial_t {\delta {\cal T}\over \delta u_t}+{\delta {\cal E}\over \delta u}
=-{\delta {\cal R}\over \delta u_t}~,
\label{varform}
\end{equation}
but now with a finite kinetic energy functional
\begin{equation}
{\cal T}={1\over 2} \epsilon^2\rho\int\! d{\bf x}\int\! d{\bf x}'
u_t({\bf x},t) {\cal G}_2({\bf x}-{\bf x}') u_t({\bf x}',t)~,
\label{kinetic_energy}
\end{equation}
and a nonlocal contribution to the dissipation function
\begin{eqnarray}
{\cal R}&=&{1\over 2} \int\! d{\bf x} u_t^2 ({\bf x},t)\nonumber \\
&&+{\epsilon\rho\over 2}\int\! d{\bf x}\int\! d{\bf x}'
u_t({\bf x},t){\cal G}_1({\bf x}-{\bf x}') u_t({\bf x}',t)~.
\label{nonlocal_dissipation}
\end{eqnarray}
The positivity of ${\cal G}_2$ renders the ``mass" of the
field dynamics positive, while there is a competition between the
signs of the local and nonlocal parts of the dissipation function.

A heuristic derivation of the leading-order changes to the contour dynamics
from the inclusion of nonzero $\epsilon$ proceeds as in Section IV.  Starting
from the dynamics for $u$ in (\ref{u_dynamics_eps2}) or
equivalently the functionals in (\ref{kinetic_energy}) and
(\ref{nonlocal_dissipation}), we recognize that
the time derivative $u_t$ and acceleration $u_{tt}$ are localized at the
domain boundary.  Using a correspondence like that in Eq. \ref{disspn_fnc_1d},
we obtain the contour representation of the kinetic energy,
\begin{equation}
{\cal T}\simeq {\rho\gamma^2\epsilon^2\over 2D}
\oint\! ds\oint\! ds'
\hat{\bf n} \cdot {\bf r}_t \ {\cal G}_2({\bf r}-{\bf r}')
\ \hat{\bf n} \cdot {\bf r}_t'~,
\label{kinetic_energy_contour}
\end{equation}
and dissipation function
\begin{eqnarray}
{\cal R}&\simeq& {\gamma\over 2D}
\oint\! ds \left(\hat{\bf n} \cdot {\bf r}_t\right)^2 \nonumber \\
&&+{\gamma^2\rho\epsilon\over 2D^2}
\oint\! ds\oint\! ds'
\hat{\bf n} \cdot {\bf r}_t{\cal G}_1({\bf r}-{\bf r}')
\hat{\bf n} \cdot {\bf r}_t'~.
\label{dissipation_contour}
\end{eqnarray}
This kind of nonlocal dissipation function occurs as well in the
analysis of certain models of solidification \cite{symmetric_model,reg_unpub},
and is conceptually similar to role of the Oseen tensor in polymer dynamics
\cite{Doi}.

The interpretation of this dynamics as a damped mechanical system is
complicated by the nonlocal nature of the kinetic energy and
dissipation functionals.
With the function ${\cal G}_1<0$, the overall rate of dissipation
is positive for small $\epsilon$ but may become negative for $\epsilon$
sufficiently large, say of order $1/\rho$.
One expects qualitatively new behavior in this limit.
Moreover, the inclusion of inertial terms in the dynamics makes the
analysis of traveling-wave states qualitatively different from that in the
overdamped limit.
Specifically, an ansatz of the form $u(x-ct)$ yields a cubic
equation for the speed $c$, with three possible roots, in contrast to the
unique speed found in the limit $\epsilon \to 0$.
These two roots most likely represent those associated with the nonequilibrium
Ising-Bloch front bifurcation discussed by Hagberg and Meron \cite{Hagberg},
and
are related to the oscillatory instabilities
discussed by Ohta, et al. \cite{Ohta} for the case of simple geometries.

Developing a theory for the asymptotic stability of the labyrinthine
patterns is one current effort \cite{Muraki}.  It is hoped that results in this
direction
will lead to a quantitative connection between the problem of finite
inhibitor diffusion ($\epsilon \ne 0$) and the onset of time-oscillatory
behavior.

\section{Nonlocal Contour Dynamics in Other Systems}
\label{sec:Nonlocal}

We have focused here on the fast-inhibitor limit of a reaction-diffusion system
and demonstrated that its behavior is well-described by a nonlocal contour
dynamics model.  In this limit, where the inhibitor is slaved to the activator,
the dynamics is a gradient flow with an energy functional of the form
\begin{equation}
{\cal E}[{\bf r}]=\Pi A + \gamma L
-{1\over 2}\rho\oint\! ds\oint\! ds' \hat{\bf t} \cdot \hat{\bf t}'
\Phi\left(R/h\right),
\label{energy_func_conclusion}
\end{equation}
with $R=\vert{\bf r}(s)-{\bf r}(s')\vert$, and a normal velocity
$U$ proportional to the force obtained variationally as
$-\hat{\bf n}\cdot \delta{\cal E}/\delta {\bf r}$, with
\begin{equation}
U= -\Pi -\gamma\kappa
+{\rho\over h}\oint\! ds'\hat {\bf R}(s,s')\times \hat{\bf t}'
\Phi'\left(R/h\right)~.
\label{normal_vel_conclusion}
\end{equation}
In this section we make specific the connections between this form of
dynamics and those found in several other quite distinct physical
systems.
These are (i) Type-I superconductors in the intermediate
state \cite{Type_Ib}, (ii) magnetic fluids in Hele-Shaw flow
\cite{Rosensweig,science,pra,pre,Cebers_integrals}, and (iii) Langmuir
monolayers of dipolar molecules \cite{jpc,Keller}.  While the
functions $\Phi$ differ from one system to the next, the common
feature we find in each is a positive bare line tension $\gamma$, and
a repulsive interaction between antiparallel tangent vectors, with
$\rho>0$.  Interfacial instabilities leading to fingered structures
are then seen to derive from a negative effective line tension arising
from the nonlocal contribution.  It is this same nonlocal coupling of
the interface that leads to the self-avoiding nature of the pattern
formation beyond the linear instability.  We discuss each of the three
systems below, pointing out the different physical origins of the
nonlocal coupling, and the different constraints on the interface
motion.

{\it Type-I Superconductors:} The intermediate state of Type-I
superconductors occurs when a thin slab of
the material below its zero-field transition temperature is placed in
a magnetic field normal to its surface \cite{Huebener}.
Rather than exhibit a complete Meissner effect, the demagnetizing
effects arising
from the sample geometry lead instead to the penetration of
the flux through the sample
in an intricate arrangement of flux domains, each of which is fingered and
often branched.   The shapes of these flux domains arise from a
competition between
the positive superconductor-normal surface energy and the
interactions between the Meissner currents which circulate at the flux domain
boundaries.   As argued many years ago \cite{Pearl,Fetter}, this self-induction
interaction retains its long-range form as in free space,
since the electromagnetic
fields in the vacuum above and below the slab are unscreened.  In the simplest
treatment of these interactions \cite{Type_Ib}, they are considered identical
to those of current loops in completely free space.  Upon averaging the
standard Coulombic self-interaction between elementary current segments
over the thickness $h$ of the slab, one finds a tangent-vector coupling for
a single interface
\begin{eqnarray}
\Phi(z)&=& -\ln\left[z^{-1}+\left(1+ z^{-2}\right)^{1/2}\right] + z -
\left[1+z^2\right]^{1/2}~,\nonumber \\
-\Phi'(z)&=& \left[1+z^{-2}\right]^{1/2}-1~,
\label{super_Psi}
\end{eqnarray}
where the characteristic length scale is $h$.  While retaining the
Coulombic form $\Phi(z)\sim 1/z$ for $z\to \infty$,  $\Phi$ has only a
logarithmic singularity at the origin, much like the Bessel function $K_0$ in
the
reaction-diffusion problem considered here.

Interface motion in superconducting systems does not conform to the simple
local-dissipation model discussed in Section IV, but rather reflects the
diffusion of the magnetic flux in the normal phase \cite{Pippard}.  Thus, the
force
(\ref{normal_vel_conclusion}) becomes a boundary condition for the
diffusion equation obeyed by the field, rather than determining the velocity
directly \cite{Type_Ia}.  In simple models of this pattern formation
\cite{Type_Ib},
in which the local dissipation model is invoked, this conservation law
leads to the generalization of the product $\Pi A$ to the derivative of a
bulk free energy density ${\cal E}_{bulk}(A)$
which arises from the competition between the field energy of the external
magnetic field and the condensation free energy of the superconducting state.
The conservation of magnetic flux is an important {\it global} constraint in
this problem, and leads to an equilibrium area fraction of the domains
that is determined primarily by the minimization of this
energy;  the shapes of the individual interfaces arise from the competing line
tension
and Biot-Savart interactions.  Unlike the reaction-diffusion problem, the
intermediate
state properties are determined fundamentally by many-domain interactions.

{\it Magnetic Fluids:} A second physical system conforming to the energetics in
Eqs.
\ref{energy_func_conclusion} and \ref{normal_vel_conclusion} is that of
thin domains of magnetic fluids \cite{Rosensweig} in the geometry of Hele-Shaw
flow.  There, the domain is trapped between two glass plates spaced a distance
$h$
apart, with the remainder of the gap filled by water.  A magnetic field applied
normal
to the plates aligns the microscopic magnetic domain in suspension, creating an
approximately uniformly magnetized droplet.  By the correspondence between
magnetization
and current-loops, the field energy associated with the domain is again
represented in terms of a self-induction interaction, with the functions $\Phi$
and $\Psi$ as in (\ref{super_Psi}) by virtue of the slab geometry.  The
amplitude $\rho$ is now proportional to $M^2h$, where $M$ is the magnetization
of the domain, and the line tension $\gamma=h\sigma$, with $\sigma$ the
ferrofluid-water surface tension.

In this hydrodynamic problem, the dynamics is well-approximated by Darcy's law
${\bf v}= -(h^2/12\eta)\bbox{\nabla} P$, where ${\bf v}$ is the $z$-averaged
in-plane
fluid velocity, $\eta$ the fluid viscosity, and $P$ a generalized pressure
including magnetic contributions \cite{science,pre}.  With the constraint of
fluid incompressibility, the pressure field is harmonic, with the
force (\ref{normal_vel_conclusion}) again acting as a boundary condition on
$P$.  The fluid incompressibility
leads directly to the conservation of the area enclosed by the boundary.

{\it Langmuir Monolayers:}
The final class of systems governed by these energetics includes amphiphilic
(Langmuir) monolayers at the air-water interface.  These consist of single-
or multi-component monomolecular films of surfactants in which the mean lateral
density is regulated externally, allowing for a study of phases and phase
transformations.  Under suitable conditions, domains of a high-density phase
appear in a background of lower density, and may be visualized by the
differential fluorescence of a dye incorporated into the layer.  One observes
various shape instabilities of these domains as conditions such as temperature
and pressure are varied \cite{McConnell_Mohwald}.  These are believed to arise
from the competing effects of line tension at the domain boundary and
long-range
electric dipole
interactions between the molecules which are oriented by the constraint of
packing.  The energetics of these interactions may be described \cite{jpc} by
taking the ``ultra-thin" limit of the magnetic formulation (\ref{super_Psi})
above,
with the cutoff $h$ being a molecular length, and the amplitude $\rho$
proportional to the square of the dipole moment density $\mu$.
In the limit of small $h$, the self-induction and Biot-Savart terms
have the form associated with infinitesimal current-carrying wires
\cite{Keller},
\begin{equation}
\Phi(z)={1\over z},\ \ \ \
-\Phi'(z)={1\over z^3}~.
\label{monolayer_Biot_Savart}
\end{equation}
A cutoff procedure must be implemented on these functions to treat
the divergences which occur when $s-s'\to 0$.  As in the
superconductor problem, in the simplest model \cite{jpc} $\Pi$ is a
Lagrange multiplier conjugate to the area.
Experiments \cite{Lee_McConnell} have shown a variety of fingered domain
shapes consistent with the boundary model.
More recent work has focused on the hydrodynamics of monolayer domains
coupled to the viscous subfluid \cite{Stone_McConnell,Lubensky_Goldstein}.

The nonlocal energy functional ${\cal E}[u]$ in
Eq. (\ref{slaved_energy}) is known also to be relevant for physical
systems quite distinct from those with dipolar interactions, appearing
for instance in models of microphase separation in block copolymers
\cite{Ohta_polymers}.  There the nonlocal coupling is long-ranged,
reflecting the connectivity of the polymers. Labyrinthine patterns
occur there as well (see also the review in
Ref. \cite{Seul_Andelman}), but are not necessarily confined to two
dimensions. It has also been remarked \cite{Weeks} that this nonlocal
interface coupling may be related to the diffusion of impurities
and/or latent heat in solidification, which leads to some degree of
interface self-avoidance seen in the development of dendrites.  Such
behavior can not be captured by purely local geometric \cite{Brower}
or boundary-layer \cite{boundary_layer} models.  Finally, observe that
the Biot-Savart coupling also appears in the contour dynamics
formulation \cite{REGDMP,Zabusky,Dritschel} of vortex patch motion in
two-dimensional ideal fluids.  Rather than exhibiting strongly
overdamped dynamics, these are of course Hamiltonian systems.

\section{Conclusions}
\label{sec:Concl}

Finally, we discuss briefly three important open issues regarding the
dynamics of labyrinthine pattern formation: generalizations to
higher spatial dimension, further elucidation of variational principles,
and derivation from microscopic chemical kinetics.

We have seen that the phenomenon of lateral inhibition necessarily
involves interactions between segments of chemical fronts that are
potentially far apart in arclength, yet close in space.  It is
precisely this nonlocality which enters the energetics
(\ref{energy_func_conclusion}) and dynamics
(\ref{normal_vel_conclusion}).  Whereas in the magnetic systems the
angular part of the current-current interactions is naturally written
as $\hat{\bf t}(s)\cdot \hat{\bf t}(s')$ in accord with the existence
of currents circulating in the direction $\hat{\bf t}$, there is no such
circulation in the reaction-diffusion problem.  But of course, as seen in
the derivation (\ref{boundary_terms}), this scalar product may equally well be
written in terms of the normal vectors as
$\hat{\bf n}(s)\cdot\hat{\bf n}(s')$.  This formulation makes it
clear that the central
issue is whether fronts oppose one another with a region of low
activator concentration in between.  The normal vector representation
allows a straightforward generalization to the interaction of
two-dimensional surfaces, for which there is no natural or unique
assignment of tangent vectors.  We suggest that certain
three-dimensional patterns may be profitably studied by models
embodying this nonlocal interaction.  In addition to the block
copolymer systems mentioned earlier, other candidates include highly
convoluted structures such brain coral, which displays labyrinthine
structures with features such as three-fold coordinated nodes like
those seen in the present work.  One may imagine that these
arise from the interplay between growth of individual members of
the colony and the competition for nutrients.

As discussed in Section \ref{sec:Memory}, a heuristically-derived
contour dynamics for small deviations from the fast-inhibitor limit
appears to conform to a rather general variational principle much like
that of a damped mechanical system.  An issue of some significance is the
extent to which the contour dynamics approach may be more rigorously
extended to incorporate the front bifurcation that ultimately occurs
for $\epsilon$ sufficiently large.  Related issues concern the
connection between such a description and spiral-wave behavior,
as well as the nature of the front bifurcation for surfaces
moving in three dimensions.

In light of the present derivation of the contour dynamics from the
FitzHugh-Nagumo model, it remains of great interest to investigate as
well whether the particular chemical kinetics \cite{Epstein}
relevant to the experiments of Lee, {\it et al.}
may be recast as an interface dynamics.
Recent work \cite{Showalter} has shown that the those very complex
kinetics have dynamics on many time scales and may be reduced to an effective
two-variable model through a sequence of slaving approximations.
While the form of that reduced description is somewhat different than the
FitzHugh-Nagumo model, the possibility that it shows similar,
near gradient-flow behavior is an intriguing area of investigation.

\section{Acknowledgments}

We are indebted to K.J. Lee, W.D. McCormick, Q. Ouyang, and H.L. Swinney
for extensive discussions concerning their experimental
work and for Fig. 1, to A.J. Bernoff, A.T. Dorsey, and K.J. Lee for detailed
comments on the manuscript, and to
D. Levermore for important suggestions at an early stage of this work.
DJM thanks Y. Kodama for introducing him to the Lie
transform.
We have also benefited from discussions with
S. Erramilli, E. Knobloch, S. Leibler, E. Meron, M.J. Shelley, and V. Hakim.
This work has been supported by an N.S.F.
Presidential Faculty Fellowship Grant DMR-9350227 and the Alfred P. Sloan
Foundation (REG), an N.S.F. Graduate Fellowship (DMP),
N.S.F. Grant DMS-9404374 and DOE Grant DE-FG02-88ER25053 (DJM).

\appendix
\section{Numerical Methods}

Here are summarized established pseudospectral methods
\cite{Canuto,Shelley_private}
that we have adapted to study both the reaction-diffusion
dynamics and the contour evolution.
For the case of a scalar partial differential equation in $1+1$ dimensions,
\begin{equation}
{\partial u\over \partial t}={\cal L}(\partial_x)u + {\cal N}[u]~,
\label{generalpde}
\end{equation}
where ${\cal L}$ is a linear operator and ${\cal N}$ is nonlinear.
We assume that the highest-order spatial derivative appears in the
linear operator.  In Fourier space (\ref{generalpde}) is
\begin{equation}
{\partial {\hat u}(k,t)\over\partial t}-\omega(k){\hat u}(k,t)=
{\hat{\cal N}}(k,t).\label{pseudosp2}
\end{equation}
with $\omega(k)={\cal L}(ik)$.
The nonlinear terms are obtained pseudospectrally
by fast Fourier transforming ${\cal N}$ in real space,
${\hat{\cal N}}(k,t)={\cal F}[{\cal N}(u(x,t))]$,
and $u(x,t)$ is computed by an inverse fast
Fourier transformation of ${\hat u}(k,t)$.
Now define
\begin{equation}
{\hat v}(k,t)={\rm e}^{-\omega(k)t}{\hat u}(k,t),\label{pdef}
\end{equation}
and multiply (\ref{pseudosp2}) by the exponential factor, yielding
\begin{equation}
{\partial {\hat v}(k,t)\over \partial t}=
{\rm e}^{-\omega(k)t}{\hat{\cal N}}\left[{\hat v}{\rm e}^{\omega(k)t}\right].
\label{pseudosp3}
\end{equation}
In the simplest Euler method, the solution to Eq. \ref{pseudosp3} is
\begin{equation}
{{\hat v}(k,t+\Delta t)-{\hat v}(k,t)\over \Delta t} \simeq
{\rm e}^{-\omega(k) t}
{\hat{\cal N}}\left({\hat v}(k,t){\rm e}^{\omega(k) t}\right),\label{euler1}
\end{equation}
which yields
\begin{equation}
{\hat u}(k,t+\Delta t) = {\rm e}^{\omega(k)\Delta t}\left[{\hat u}(k,t) +
\Delta t {\hat{\cal N}}(k,t)\right]~.
\label{finaleuler}
\end{equation}
The exponentiation of the growth rate $\omega(k)$ in (\ref{finaleuler})
plays a useful role in guaranteeing
stability for a diffusive linear operator ($\omega(k)=-\gamma k^2$).
The usual stability considerations \cite{Numerical_Recipes} would require
a time step $\Delta t$ such that
for large values of momentum (near the Brillouin zone edge
$k_{{\rm max}}=\pi/a$, with $a$ the lattice spacing in real space)
the quantity $k^2\Delta t$ be less than unity.  This requires an extremely
small
time step, rendering the calculation prohibitively slow.
Here, even if $k_{{\rm max}}^2 \Delta t >1$, the calculation is stable due to
the
incorporation of the exact dynamics of the linear operator, namely the
exponential damping at high momentum.

The generalization of Eq. \ref{finaleuler} to a $4$th order Runge-Kutta method
proceeds as follows.  Define
\begin{equation}
L_{1/2}(k)={\rm e}^{\omega(k)\Delta t/2}, \ \ \ \
L(k)={\rm e}^{\omega(k)\Delta t}~,
\label{Ldefine}
\end{equation}
and the intermediate results
\begin{eqnarray}
{\hat U}_1(k)&=&L_{1/2}(k)\left[{\hat u}+{1\over 2}\Delta t
{\hat{\cal N}}\left({\hat u}\right)\right]\\
{\hat U}_2(k)&=&L_{1/2}(k){\hat u}+{1\over 2}\Delta t {\hat{\cal N}}_1\\
{\hat U}_3(k)&=&L(k){\hat u}+\Delta tL_{1/2}(k){\hat{\cal N}}_2~,
\end{eqnarray}
with ${\hat{\cal N}}_j={\hat{\cal N}}\left(\hat U_j\right)$.
Then the time-stepping routine analogous to (\ref{finaleuler}) is
\begin{eqnarray}
{\hat u}\left(t+\Delta t\right)&=& L(k){\hat u}
+{1\over 6} \Delta t L(k){\hat{\cal N}}
+{1\over 3} \Delta t L_{1/2}(k){\hat{\cal N}}_1 \nonumber \\
&&+ {1\over 3} \Delta t L_{1/2}(k){\hat{\cal N}}_2 + {1\over 6} \Delta t
{\hat{\cal N}}_3~.
\label{scaler_Runge_Kutta}
\end{eqnarray}

To generalize this method to situations with $n$ coupled
variables $u_i$ ($i=1,2,\ldots,n$),
we write the equation of motion in Fourier space in vectorial form
\begin{equation}
{\hat{\bf u}}_t={\bf \Omega}\cdot {\hat{\bf u}} +
\hat{\bf {\cal N}},
\label{manyv1}
\end{equation}
with ${\bf \Omega}$ a matrix of wavevector-dependent growth
rates in Fourier space
and $\hat{\bf {\cal N}}$ a vector of nonlinear terms obtained pseudospectrally.
Now suppose that $\det{{\bf \Omega}-\omega {\bf I}}=0$
has as solutions eigenvalues $\omega_i$
and associated eigenvectors $\hat{\bf e}_i$ ($i=1,\ldots,n$).
The matrix ${\bf T}$ whose columns are the components of the eigenvectors
defines a linear transformation between ${\bf u}$ and the vector ${\bf a}$
of expansion coefficients in the basis of eigenvectors.  That is,
${\hat{\bf u}} ={\bf T}\cdot{\bf a}$,
where ${\bf T}^{-1}\cdot {\bf T}={\bf I}$.
The equation of motion (\ref{manyv1}) then becomes
\begin{equation}
{\bf a}_t={\bf T}^{-1}\cdot
{\bf \Omega}\cdot {\bf T}\cdot {\bf a} + {\bf T}^{-1}\cdot \hat{\bf {\cal N}}.
\label{neweom1}
\end{equation}
The matrix ${\bf D}\equiv {\bf T}^{-1}\cdot {\bf \Omega}\cdot {\bf T}$ is
diagonal.
Now let ${\bf v}\equiv  \exp(-{\bf D}t) {\bf a}$, so
\begin{equation}
{\bf v}_t={\rm e}^{-{\bf D}t} {\bf T}^{-1}\cdot \hat{\bf {\cal N}}~.
\label{betadef1}
\end{equation}
It follows that for an Euler method, one need only know the matrix
\begin{equation}
{\bf L}\equiv {\bf T}\cdot {\rm e}^{{\bf D}\Delta t}{\bf T}^{-1}~,
\label{betadef2}
\end{equation}
for then
\begin{equation}
{\hat{\bf u}}(k,t+\Delta t) = {\bf L}\cdot\left[{\hat{\bf u}}(k,t) +
 \Delta t \hat{\bf {\cal N}}(k,t)\right]~.
\label{eulergeneral}
\end{equation}
whereas in the Runge-Kutta method we obtain the vectorial analog of Eq.
\ref{scaler_Runge_Kutta}
\begin{eqnarray}
{\hat{\bf u}}\left(k,t+\Delta t\right)&=& {\bf L}\cdot{\hat{\bf u}}(k,t)
+{1\over 6} \Delta t {\bf L}\cdot\hat{\bf {\cal N}}
+{1\over 3} \Delta t {\bf L}_{1/2}\cdot\hat{\bf {\cal N}}_1 \nonumber \\
&&+ {1\over 3} \Delta t {\bf L}_{1/2}\cdot\hat{\bf {\cal N}}_2
+ {1\over 6} \Delta t \hat{\bf {\cal N}}_3~.
\end{eqnarray}
where
\begin{equation}
{\bf L}_{1/2}\equiv {\bf T}\cdot {\rm e}^{{\bf D}\Delta t/2}{\bf T}^{-1}~,
\label{betadef3}
\end{equation}
and the intermediate results are
\begin{eqnarray}
\hat{\bf u}_1&=&{\bf L}_{1/2}\cdot\left[{\hat{\bf u}}+{1\over 2}\Delta t
\hat{\bf {\cal N}}\left({\hat{\bf u}}\right)\right]\\
\hat{\bf u}_2&=&{\bf L}_{1/2}\cdot{\hat{\bf u}}+{1\over 2}\Delta t \hat{\bf
{\cal N}}_1\\
\hat{\bf u}_3&=&{\bf L}\cdot{\hat{\bf u}}+\Delta t {\bf L}_{1/2}\cdot\hat{\bf
{\cal N}}_2~,
\label{Uvecdefine}
\end{eqnarray}
with $\hat{\bf {\cal N}}_j=\hat{\bf {\cal N}}\left(\hat{\bf u}_j\right)$.

{\it The fast-inhibitor limit:}
In the limit $\epsilon=0$, Eq. (\ref{slave1}), the single equation of
motion for $u$ contains a contribution that while nonlocal is
nevertheless linear.  Since it is a convolution, it is local in
Fourier space and can be incorporated directly into the
linear operator ${\cal L}$.  The resulting transform is
\begin{equation}
\omega(k)=-Dk^2-r+\rho{k^2 \over 1+k^2}~,
\label{omega_fast}
\end{equation}
precisely the growth rate of the mode $+$ in Eq. \ref{growths}.
The diffusive contribution $-Dk^2$ dominates at large wavevector.

{\it Contour dynamics}: In numerical studies of the contour dynamics
we have employed techniques described elsewhere \cite{pra,jpc}, summarized
briefly here.  Starting from an equation of motion in the form
\begin{equation}
{\bf r}_t=U\hat{\bf n}+W\hat{\bf t}~,
\label{generaleom}
\end{equation}
we study the evolution of the tangent angle \cite{Brower}
$\theta(s)$, related to the curvature
by $\kappa(s)=\partial \theta/\partial s$,
\cite{Brower}
\begin{equation}
{\partial\theta\over\partial t} =
 - {\partial U\over\partial s}+\kappa W~.
\label{thetaeom}
\end{equation}
The spectral method described above requires that we utilize a
periodic function.  A convenient choice is the deviation $\psi$ of $\theta$
from the linear form $\theta=2\pi s/L$ for a circle,
\begin{equation}
\theta(\alpha,t)=2\pi\alpha+\psi(\alpha,t)~.
\label{psi_defn}
\end{equation}
A natural choice of gauge is that of ``relative arclength," for which
equally spaced points in the parametrization $\alpha=s/L$ remain
equally spaced in time.  This corresponds to a tangential velocity
\cite{pra,jpc,Brower}
\begin{equation}
W(\alpha)=L\left(\alpha \int_0^1\! d\alpha'\, \kappa U
-\int_0^\alpha\! d\alpha'\kappa U\right)~.
\label{Weqn}
\end{equation}
The highest-order arclength derivative in the $\psi$ dynamics is then
diffusive,
\begin{equation}
{\partial \psi\over\partial t} ={\gamma\over L^2}{\partial^2 \psi
\over \partial \alpha^2} + \cdots~,
\label{highest_deriv}
\end{equation}
and amenable to the integrating factor method outlined above.
The dynamical variables of the problem are then $\psi$ and the
contour length $L$ which obeys the simple evolution
\begin{equation}
L_t = \oint\! ds \kappa U~.
\label{L_evolution}
\end{equation}

\begin{figure}
\caption{Chemical pattern formation in the iodide-fer-\break rocyanide-sulfite
reaction of Lee, McCormick, Ouyang, and Swinney \protect{\cite{Lee}}.
Low and high pH regions appear respectively as white and black by means of
a pH indicator.  Times
proceed from upper left to lower right in hours
following a perturbation. Figure courtesy of Lee, {\it et al.}, and
reproduced from Ref \protect{\cite{prl}}.
\label{expt_fig}}
\end{figure}

\begin{figure}
\caption{Simulation of the reaction-diffusion PDEs showing a compact
initial condition undergoing a fingering instability, eventually
producing a space-filling labyrinth.  Panels are contour plots with
$u\le 0.5$ shown white, and $u\ge 0.5$ shown black, at rescaled times
of $\tau=0,5,10,15,20,25$.  The
parameters in Eq. \protect{\ref{uvequations2a}} are $D=0.01$, $\rho=0.15$,
and $r=0.52$; thus, the state $u=1$ (black) is less stable than
$u=0$ (white).
\label{intro_sim_fig1}}
\end{figure}

\begin{figure}
\caption{Reaction-diffusion simulations as in
Fig. \protect{\ref{intro_sim_fig1}}, but starting with a branched
domain, and with $D=0.01$, $\rho=0.10$, and $r=0.60$.  With these
parameter values, a circular localized state is the stable
configuration to which the system relaxes.
\label{intro_sim_fig2}}
\end{figure}

\begin{figure}
\caption{Regions of the $r-\rho$ parameter space in which Turing
bifurcations are possible (above solid lines) and forbidden (below)
within linear stability.  Inset shows growth rate of the most
dangerous mode as a function of wavenumber both in the linearly stable
and unstable regions, for values indicated by the symbols.
\label{linstab_fig}}
\end{figure}

\begin{figure}
\caption{Space-time portrait of the formation of a localized state in
one dimension by the shrinkage of a region of $u=1$.  Activator is
shown as solid line, inhibitor as dashed line, with time increasing
downwards.  Solid symbols locate the value $u=1/2$.  Parameter values
are: $D=0.01$, $r=0.55$, $\rho=0.15$.
\label{localized_state_fig}}
\end{figure}

\begin{figure}
\caption{Space-time portrait of the growth of a localized state in
one dimension from a small nucleated domain.  Activator is shown as solid line,
inhibitor as dashed line, with time increasing downwards.  Parameters are:
$D=0.01$, $r=0.52$, $\rho=0.15$. \label{localized_state_fig2}}
\end{figure}

\begin{figure}
\caption{Front dynamics.  (a) Velocity of a single front as a function of the
activator-inhibitor-coupling, for diffusion constant $D=0.0001$,
normalized by the velocity for $\rho=0$. Symbols indicate results of
numerical simulations of the reaction-diffusion equations, while solid line
are predictions from the asymptotic analysis with the corrected friction
coefficient given by Eq. \protect{\ref{one_front_Q_more}}.
(b) Asymptotics for separation versus time for two approaching fronts in
one dimension, compared with simulation results for $D=0.0001$, $r=0.515$,
and  $\rho=0.02$.
Dashed line is prediction of asymptotics with bare friction coefficient, while
the dotted line includes order $\rho$ correction to front profile and compares
well with the full numerical solution to the reaction-diffusion equations
(solid).
Inset (c) shows the equilibrium separation of the fronts versus $\rho$ for both
simulation (solid symbols) and asymptotics (solid line) from Eq.
\protect{\ref{L_equil}}.\label{front_dynamics_fig}}
\end{figure}

\begin{figure}
\caption{Illustration of the Biot-Savart interactions of nearby chemical
fronts, as described by the nonlocal contribution to the force law in
Eq. \protect{\ref{normal_velocity}}.\label{Biot_Savart_fig}}
\end{figure}

\begin{figure}
\caption{Regions of existence and marginal stability curves of
circular localized states in a rescaled parameter space, as deduced
from the equilibrium condition of Eq. \protect{\ref{disc_radius}} and
the linear stability result of
Eq. \protect{\ref{azi_exp}}.\label{circle_fig}}
\end{figure}

\begin{figure}
\caption{Energy of circular domains as a function of radius, from Eq.
\protect{\ref{circle_energy}}, for various values of the
activator-inhibitor coupling $\rho$, with $D=0.01$, and $r=0.6$.
\label{circle_energy_fig}}
\end{figure}

\begin{figure}
\caption{Mechanism of instability of a straight chemical front.  The
accumulation of inhibitor in the neighborhood of segments of the
interface that are concave inwards leads to the deepening of the those
deformations, and vice-versa for regions that are convex outwards.
\label{front_instability_fig}}
\end{figure}

\begin{figure}
\caption{Stability diagram for stripes, in the asymptotically rescaled
parameter space.  The coalescence of the sinuous instability of a
stripe and the instability of a single front as $r\to 1/2$ reflects
the diverging equilibrium width of the stripe.  Note that the
sinuous instability lies in the region of azimuthally-stable
localized states.
\label{stripe_stability_fig}}
\end{figure}

\begin{figure}
\caption{Spectrum of growth rates for instabilities of isolated planar
fronts, and the sinuous and varicose instabilities of stripes of
finite width.  Parameters are $D=0.01$, $\rho=0.3$, and $2Q^*=1.5$.
The single front is more unstable than the finite-width stripe, while
the varicose mode is the most stable, particularly as $k\to 0$.
\label{front_stability_fig}}
\end{figure}

\begin{figure}
\caption{Contour plots of the activator $u$ from numerical simulations
of the reaction-diffusion model in two dimensions.  Parameters are
$D=0.01$, $r=0.40$, and $\rho=0.20$.  In analogy with the
one-dimensional front-repulsion phenomenon shown in
Fig. \protect{\ref{localized_state_fig}}, the approaching chemical
fronts do not cross, and equilibrate to a simple stripe.
Rescaled times are in increments of $\Delta\tau=6$ from upper left to lower
right.
\label{2dfronts_fig}}
\end{figure}

\begin{figure}
\caption{Pattern formation with multiple domains of activator.
$D=0.01$, $r=0.60$, $\rho=0.30$, $\Delta\tau=2.36$.
The domains each undergo fingering instabilities, but do not merge.
\label{multiple_domains_fig}}
\end{figure}

\begin{figure}
\caption{Numerical simulation of contour dynamics showing
destabilization of a circle. Lower panel shows perimeter as a function
of time, illustrating the asymptotically linear growth.  Parameters
are: $\tilde r=0.212$, $\tilde \rho=0.796$, with $\Delta\tau=25$.
\label{circle_to_stripe_fig}}
\end{figure}

\begin{figure}
\caption{Tip-splitting in the contour dynamics.  Initial condition is
a localized state perturbed by a small $n=2$ distortion.
Parameters are  $\tilde r=0.212$, $\tilde \rho=0.90$, $\Delta \tau=10$;
both the $n=2$ and $n=3$ modes are unstable.
\label{tip_splitting_fig}}
\end{figure}

\begin{figure}
\caption{Instability of a $4$-fold vertex.  Contour dynamics results,
the initial condition being a $4$-fold perturbed localized state with
an additional very small $n=2$ distortion.  $\tilde r=0.212$, $\tilde
\rho=1.10$,
$\Delta\tau=5$.
\label{vertex_fig}}
\end{figure}

\begin{figure}
\caption{Growth of a labyrinth from a compact initial condition.  Contour
dynamics results proceed in time from upper left to lower right, spaced by
$\Delta \tau=5$.  Parameters are $\tilde r=0.21$, $\tilde \rho=1.10$, and
initial condition is a distorted circle of radius $R=6.0$.
\label{circle_to_labyrinth_fig}}
\end{figure}

\begin{figure}
\caption{Relaxation of a labyrinthine pattern to a localized state.  Initial
condition is close to the final panel in Fig.
\protect{\ref{circle_to_labyrinth_fig}},
with parameters $\tilde r=0.21$, $\tilde \rho=0.601$, and time between figures
of $\Delta \tau=25$.
\label{labyrinth_relax_fig}}
\end{figure}


\begin{references}

\bibitem[**]{present} Present address.

\bibitem{Lee} K.J. Lee, W.D. McCormick, Q. Ouyang, and H.L. Swinney,
Science {\bf 261}, 192 (1993).

\bibitem{Lee_long} K.J. Lee and H.L. Swinney, Phys. Rev. E {\bf 51}, 1899
(1995).

\bibitem{Turing} A.M. Turing, Philos. Trans. R. Soc. London Ser. B
{\bf 237}, 37 (1952).

\bibitem{Boissonade} V. Castets, E. Dulos, J. Boissonade, and P. De Kepper,
Phys. Rev. Lett. {\bf 64}, 2953 (1990).

\bibitem{Swinney} Q. Ouyang and H.L. Swinney, Nature {\bf 352},
610 (1991).

\bibitem{Turing_review} For a review of recent experiments on the Turing
instability, see I. Lengyel and I.R. Epstein, Acc. Chem. Res. {\bf 26},
235 (1993).

\bibitem{Koga} S. Koga and Y. Kuramoto, Prog. Theor. Phys. {\bf 63},
106 (1980).

\bibitem{Ohta} T. Ohta, M. Mimura, and R. Kobayashi, Physica D
{\bf 34}, 115 (1989).

\bibitem{Knobloch} Localized states may also under oscillatory instabilities
\protect{\cite{Ohta}}.  The interplay between these and the steady
fingering instabilities has been considered by P. Hirschberg, V. Kirk,
and E. Knobloch, Phys. Lett. A {\bf 172}, 141 (1992).

\bibitem{prl} D.M. Petrich and R.E. Goldstein, Phys. Rev.
Lett. {\bf 72}, 1120 (1994).

\bibitem{FitzHugh} R. FitzHugh, Biophys. J. {\bf 1}, 445 (1961); J.S. Nagumo,
S. Arimoto, and Y.
Yoshizawa, Proc. IRE {\bf 50}, 2061 (1962); R. FitzHugh, in {\it Biological
Engineering}, ed.
H.P. Schwan (McGraw-Hill, New York, 1969);

\bibitem{Murray} J.D. Murray, {\it Mathematical Biology\/} (Springer-Verlag,
New York, 1989).

\bibitem{JKeller} J. Rubinstein, P. Sternberg, and J. Keller,
SIAM J. Appl. Math. {\bf 49}, 116 (1989).

\bibitem{Meinhardt} A. Gierer and H. Meinhardt, in {\it Lectures on
Mathematics in the Life Sciences}, vol. 7 (American Mathematical
Society, Providence, RI, 1974), p. 163; H. Meinhardt, Differentiation
{\bf 6}, 117 (1976).

\bibitem{Ermentrout} G.B. Ermentrout, S.P. Hastings, and W.C. Troy,
SIAM J. Appl. Math {\bf 44}, 1133 (1984).

\bibitem{McConnell_Mohwald} For reviews, see H. M\"ohwald, Annu. Rev. Phys.
Chem. {\bf 41}, 441 (1990); H.M. McConnell, {\it ibid.} {\bf 42}, 171
(1991).

\bibitem{Rosensweig} R.E. Rosensweig, {\it Ferrohydrodynamics\/}
(Cambridge University Press, Cambridge, 1985.

\bibitem{Cebers} A.O. Tsebers and M.M. Mairov, Magnetohydrodynamics
{\bf 16}, 21 (1980).

\bibitem{Boudouvis} A.G. Boudouvis, J.L. Puchalla, and L.E. Scriven, J.
Colloid Interface Sci. {\bf 124}, 688 (1988).

\bibitem{science}
A.J. Dickstein, S. Erramilli, R.E. Goldstein, D.P. Jackson,
and S.A. Langer, Science {\bf 261}, 1012 (1993).

\bibitem{Huebener} R.P. Huebener, {\it Magnetic Flux Structures in
Superconductors\/} (Springer-Verlag, New York, 1979).

\bibitem{Seul_films} M.~Seul, L.R.~Monar, L.~O'Gorman, and R.~Wolfe, Science
{\bf 254}, 1616 (1991).

\bibitem{Seul_Andelman} M. Seul and D. Andelman, Science {\bf 267}, 476 (1995).

\bibitem{Gray_Scott} P. Gray and S.K. Scott, Chem. Eng. Sci.
{\bf 38}, 29 (1983); {\it ibid.\/} {\bf 39}, 1087 (1984); J.
Phys. Chem. {\bf 89}, 22 (1985).

\bibitem{Pearson} J.E. Pearson, Science {\bf 261}, 189 (1993).

\bibitem{Meron} A. Hagberg and E. Meron, Phys. Rev. Lett.
{\bf 72}, 2494 (1994); Nonlinearity {\bf 7}, 805 (1994).

\bibitem{Hagberg} A. Hagberg and E. Meron, Chaos {\bf 4}, 477
(1994); C. Elphick, A. Hagberg, and E. Meron, Phys. Rev. E {\bf 51},
3052 (1995).

\bibitem{Lee2} K.J. Lee, W.D. McCormick, J.E. Pearson, and H.L. Swinney,
Nature {\bf 214}, 215 (1994).

\bibitem{Reynolds} W.N. Reynolds, J.E. Pearson, and S. Ponce-Dawson,
Phys. Rev. Lett. {\bf 72}, 2797 (1994).

\bibitem{Krischer} K. Krischer and A. Mikhailov, Phys. Rev. Lett. {\bf 73},
3165 (1994).

\bibitem{Fife} See, e.g. P.C. Fife, {\it Dynamics of Internal Layers and
Diffusive Interfaces\/} (SIAM, Philadelphia, 1988), and references therein.

\bibitem{spiral_waves} J.P. Keener, SIAM J. Appl. Math. {\bf 46},
1039 (1986); A.T. Winfree, SIAM Review {\bf 32}, 1 (1990); A.J. Bernoff,
Physica D {\bf 53}, 125 (1991);
E. Meron, Phys. Reports {\bf 218}, 1 (1992), and references therein.

\bibitem{Bernoff_private} We are grateful to A.J. Bernoff, (private
communication,
1993) for discussions concerning this point.

\bibitem{Ohta2} T. Ohta, A. Ito, and A. Tetsuka, Phys. Rev. A
{\bf 42}, 3225 (1990).

\bibitem{Swift_Hohenberg} J.B. Swift and P.C. Hohenberg, Phys.
Rev. A {\bf 15}, 319 (1977).

\bibitem{Lifshitz} R.M. Hornreich, M. Luban, and S. Shtrikman,
Phys. Rev. Lett. {\bf 35}, 1678 (1975).

\bibitem{Goldstein} H. Goldstein, {\it Classical Mechanics} 2d ed.
(Addison-Wesley,
Reading, 1980), p. 24.

\bibitem{Type_Ia} A.T. Dorsey, Annals of Physics {\bf 233}, 248 (1994).

\bibitem{pra} S.A. Langer, R.E. Goldstein, and D.P. Jackson, Phys.
Rev. A {\bf 46}, 4894 (1992).

\bibitem{Carrier} G.F. Carrier, M. Krook, and C.E. Pearson, {\it Functions
of a Complex Variable\/} (McGraw-Hill, New York, 1966).

\bibitem{Mullins_Sekerka} W.W. Mullins and R.F. Sekerka, J. Appl. Phys.
{\bf 35}, 444 (1964).

\bibitem{pre} D.P. Jackson, R.E. Goldstein, and A.O. Cebers, Phys. Rev.
E {\bf 50}, 298 (1994).

\bibitem{jpc} R.E. Goldstein and D.P. Jackson, J. Phys. Chem.
{\bf 98}, 9626 (1994).

\bibitem{Duplantier} B. Duplantier, R.E. Goldstein, V. Romero-Roch\'{\i}n,
and A.I. Pesci, Phys. Rev. Lett. {\bf 65}, 508 (1990).

\bibitem{Verwey} E.J.W. Verwey and J.Th.G. Overbeek, {\it Theory of the
Stability
of Lyophobic Colloids\/} (Elsevier, Amsterdam, 1948).

\bibitem{Landau_elasticity} L.D. Landau and E.M. Lifshitz, {\it Theory of
Elasticity} (Pergamon Press, New York, 1959), \S 17,18.

\bibitem{Griffiths} J. Langer and D.A. Singer, Topology {\bf 24}, 75 (1985).

\bibitem{folding} R.E. Goldstein and S.A. Langer, Phys. Rev. Lett.
{\bf 75}, 1094 (1995).

\bibitem{Brower} R.C. Brower, D.A. Kessler, J. Koplik, and H. Levine,
Phys. Rev. A {\bf 29}, 1335 (1984).

\bibitem{Steiner} See for instance the discussion of Steiner trees
in C. Isenberg, {\it The Science of Soap Films and Soap Bubbles\/}
(Dover, New York, 1992), Chap. 3.

\bibitem{symmetric_model} J.S. Langer, Acta Metal. {\bf 25},
1121 (1977).

\bibitem{reg_unpub} R.E. Goldstein, ``Variational Structure of a Model
for Driven Interface Motion," unpublished.

\bibitem{Doi} M. Doi and S.F. Edwards, {\it The Theory of Polymer Dynamics\/}
(Oxford University Press, New York, 1986).

\bibitem{Muraki} D.J. Muraki and R.E. Goldstein, in preparation.

\bibitem{Type_Ib} R.E. Goldstein, D.P. Jackson, and A.T. Dorsey,
``Cur-\break rent-Loop Model for the Intermediate State of Type-I
Superconductors,"
preprint (1996).

\bibitem{Cebers_integrals} A.O. Cebers, Magnetohydrodynamics {\bf 25}, 149
(1989).

\bibitem{Keller} D.J. Keller, J.P. Korb, and H.M. McConnell, J. Phys.
Chem. {\bf 91}, 6417 (1987).

\bibitem{Pearl} J. Pearl, Appl. Phys. Lett. {\bf 5}, 65 (1964).

\bibitem{Fetter} A.L. Fetter and P.C. Hohenberg, Phys. Rev.
{\bf 159}, 330 (1967).

\bibitem{Pippard} A.B. Pippard, Phil. Mag {\bf 41}, 243 (1950).

\bibitem{Lee_McConnell} K.Y.C. Lee and H.M. McConnell, J. Phys. Chem.
{\bf 97}, 9532 (1993).

\bibitem{Stone_McConnell} H.A. Stone and H.M. McConnell, Proc. Roy. Soc. Lond.
A {\bf 448}, 97 (1995).

\bibitem{Lubensky_Goldstein} D.K. Lubensky and R.E. Goldstein, ``Hydrodynamics
of Monolayer Domains at the Air-Water Interface," Phys. of Fluids, in
press (1996).

\bibitem{Ohta_polymers} T. Ohta and K. Kawasaki, Macromolecules {\bf 19},
2621 (1986).

\bibitem{Weeks} J. Weeks, private communication (1992).

\bibitem{boundary_layer} E. Ben-Jacob, N. Goldenfeld, J.S. Langer,
and G. Sch\"on, Phys. Rev. A {\bf 29}, 330 (1984).

\bibitem{REGDMP} R.E. Goldstein and D.M. Petrich, Phys. Rev.
Lett. {\bf 69}, 555 (1992).

\bibitem{Zabusky} N.J. Zabusky, M.H. Hughes, and K.V. Roberts, J.
Comput. Phys. {\bf 30}, 96 (1976).

\bibitem{Dritschel} D.G. Dritschel, J. Fluid Mech. {\bf 172},
 157 (1986).

\bibitem{Epstein} E.C. Edblom, M. Gy\"orgyi, M. Orb\'an, and I.R. Epstein,
J. Am. Chem. Soc. {\bf 108}, 2826 (1986).

\bibitem{Showalter} V. G\'asp\'ar and K. Showalter,
J. Phys. Chem. {\bf 94}, 4973 (1990), and references therein.

\bibitem{Canuto} See e.g., {\it Spectral Methods in Fluid Mechanics},
C. Canuto, M.Y. Hussaini, A. Quarteroni, and T.A. Zang, eds. (New York,
Springer-Verlag, 1988), p. 112; R.S. Rogallo, NASA TM-73203 (1977); P.R.
Spalart, NASA TM-88222 (1986);

\bibitem{Shelley_private} We are indebted to M.J. Shelley for valuable
discussions
regarding these numerical methods.

\bibitem{Numerical_Recipes} W.H. Press, S.A. Teukolsky, W.T. Vetterling, and
B.P. Flannery, {\it Numerical Recipes in C\/}, 2nd ed. (Cambridge University
Press,
Cambridge, 1992).

\end{references}
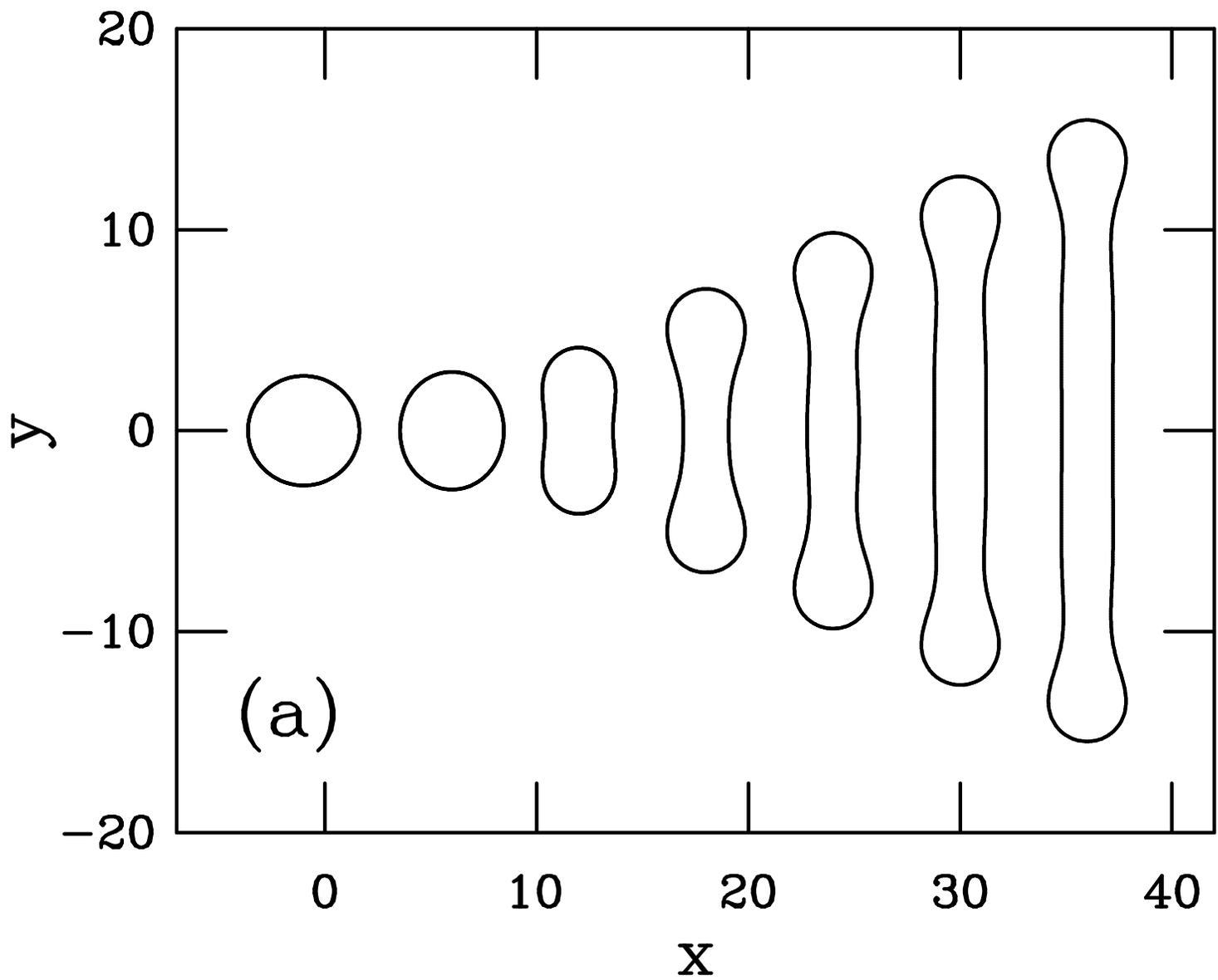
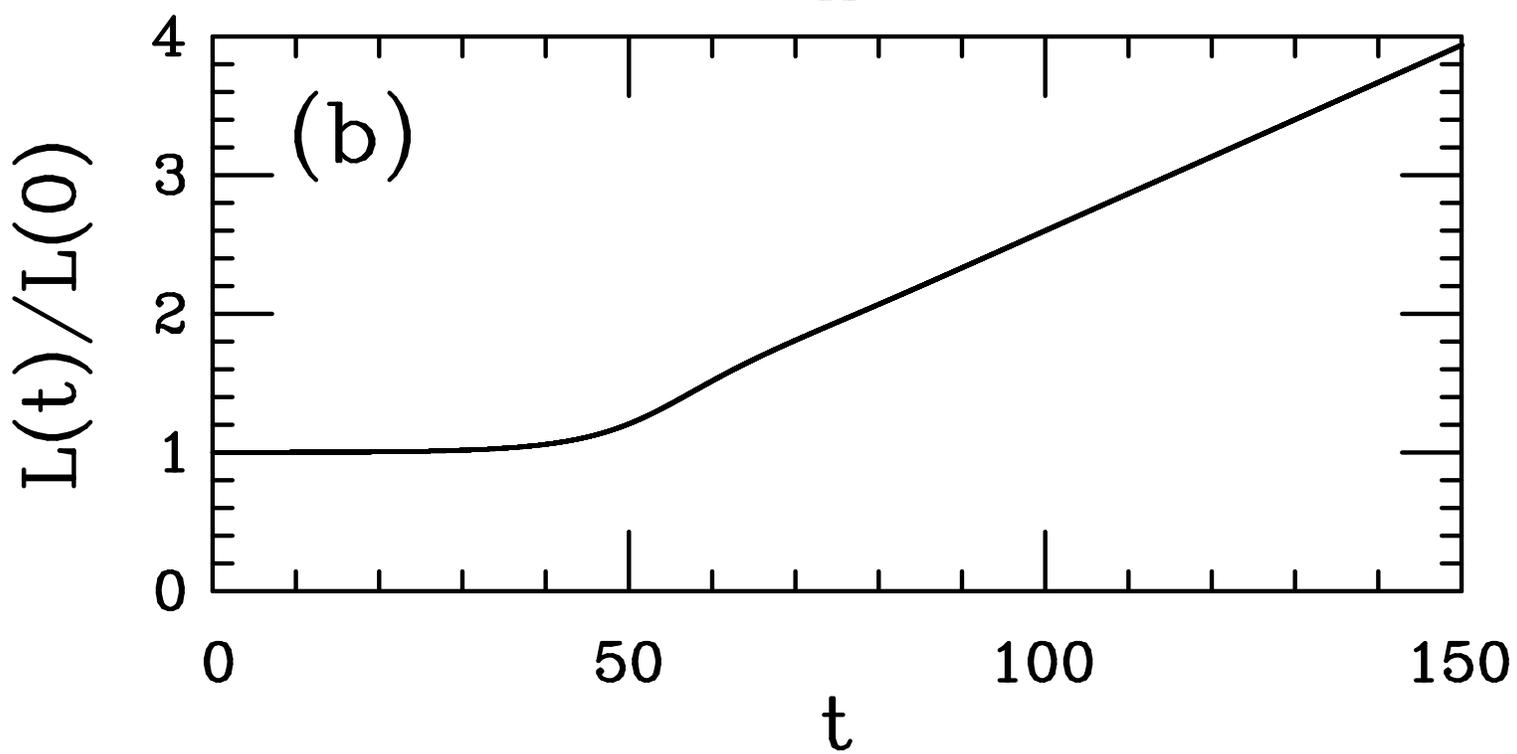
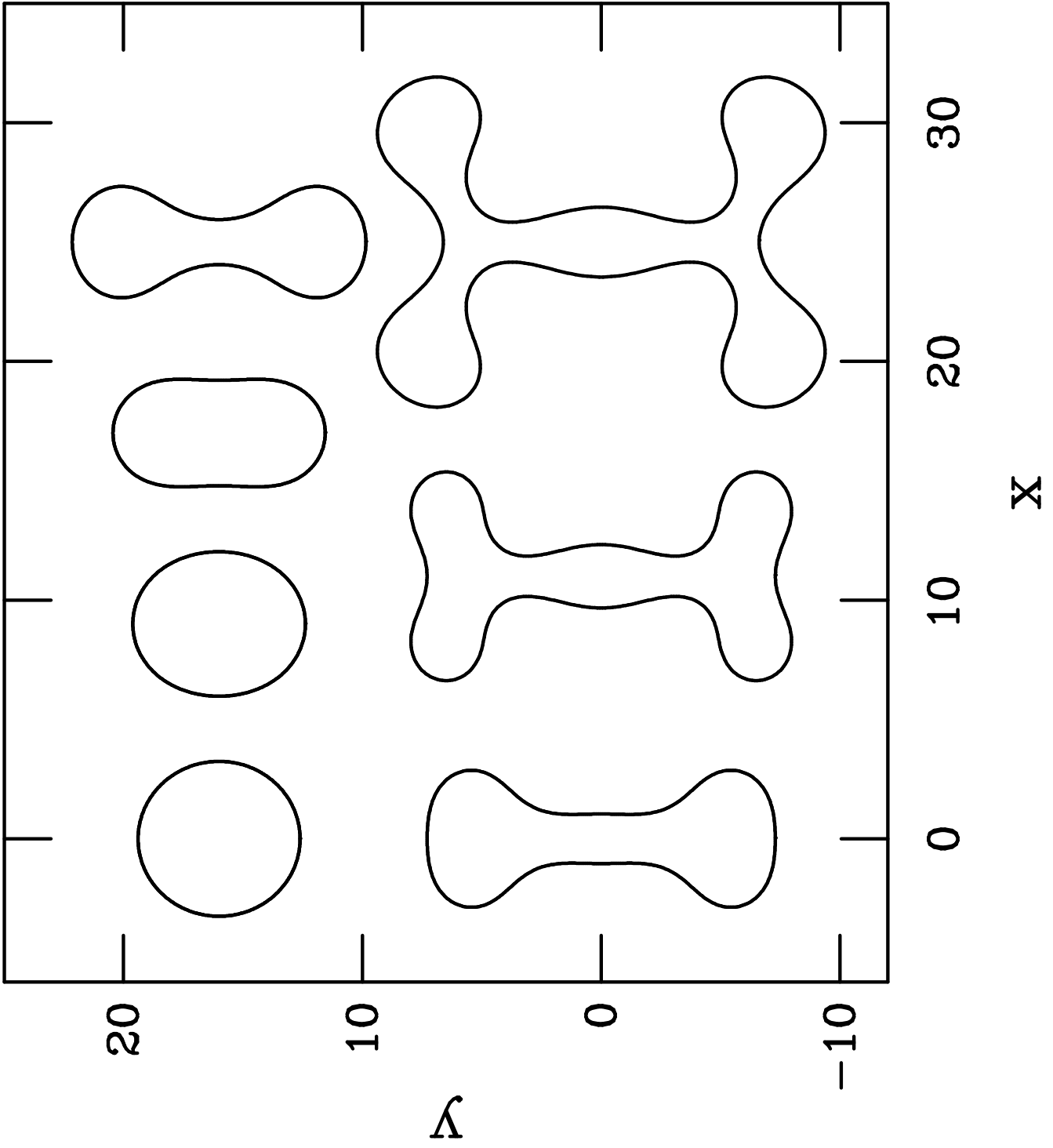
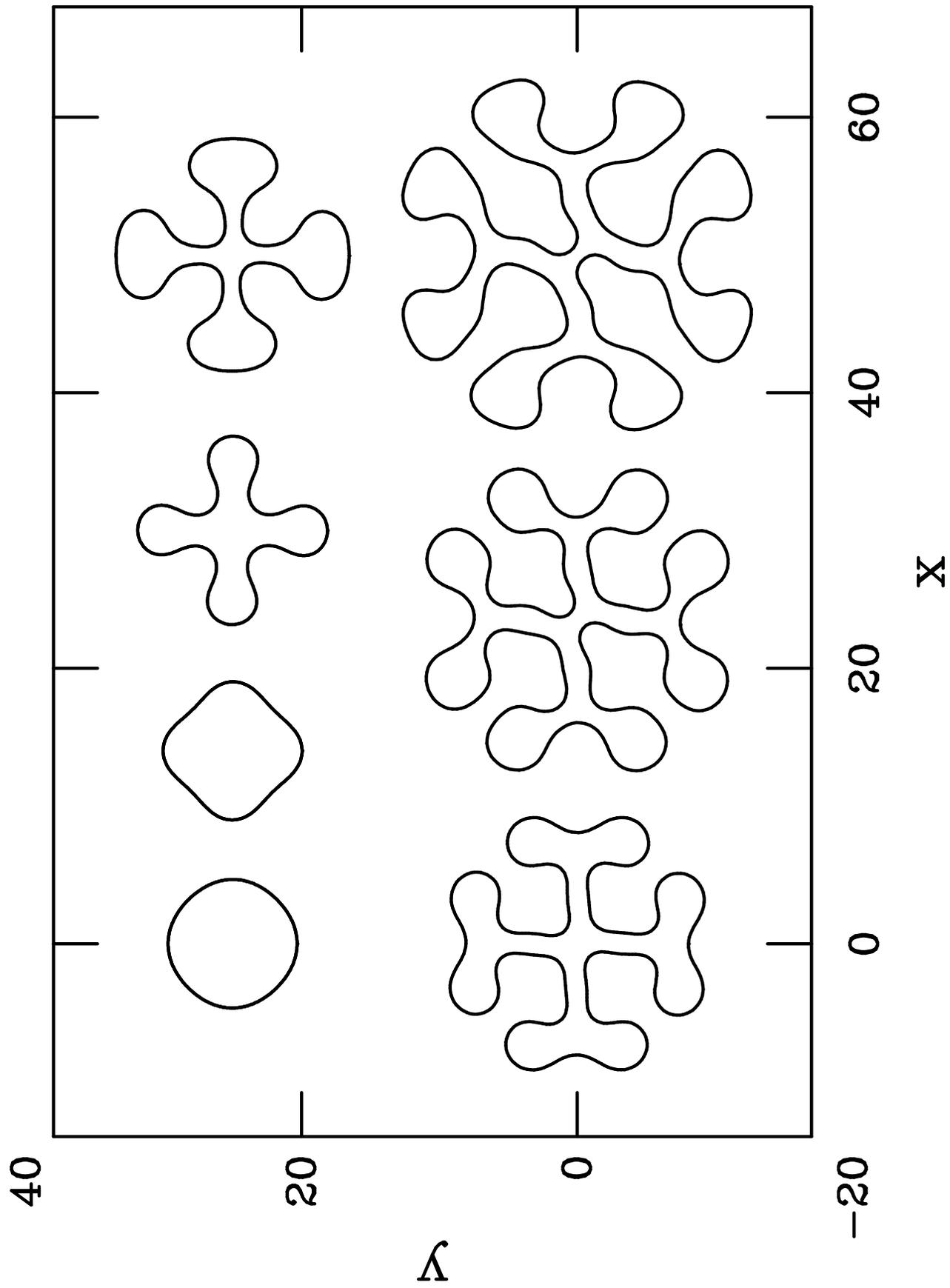
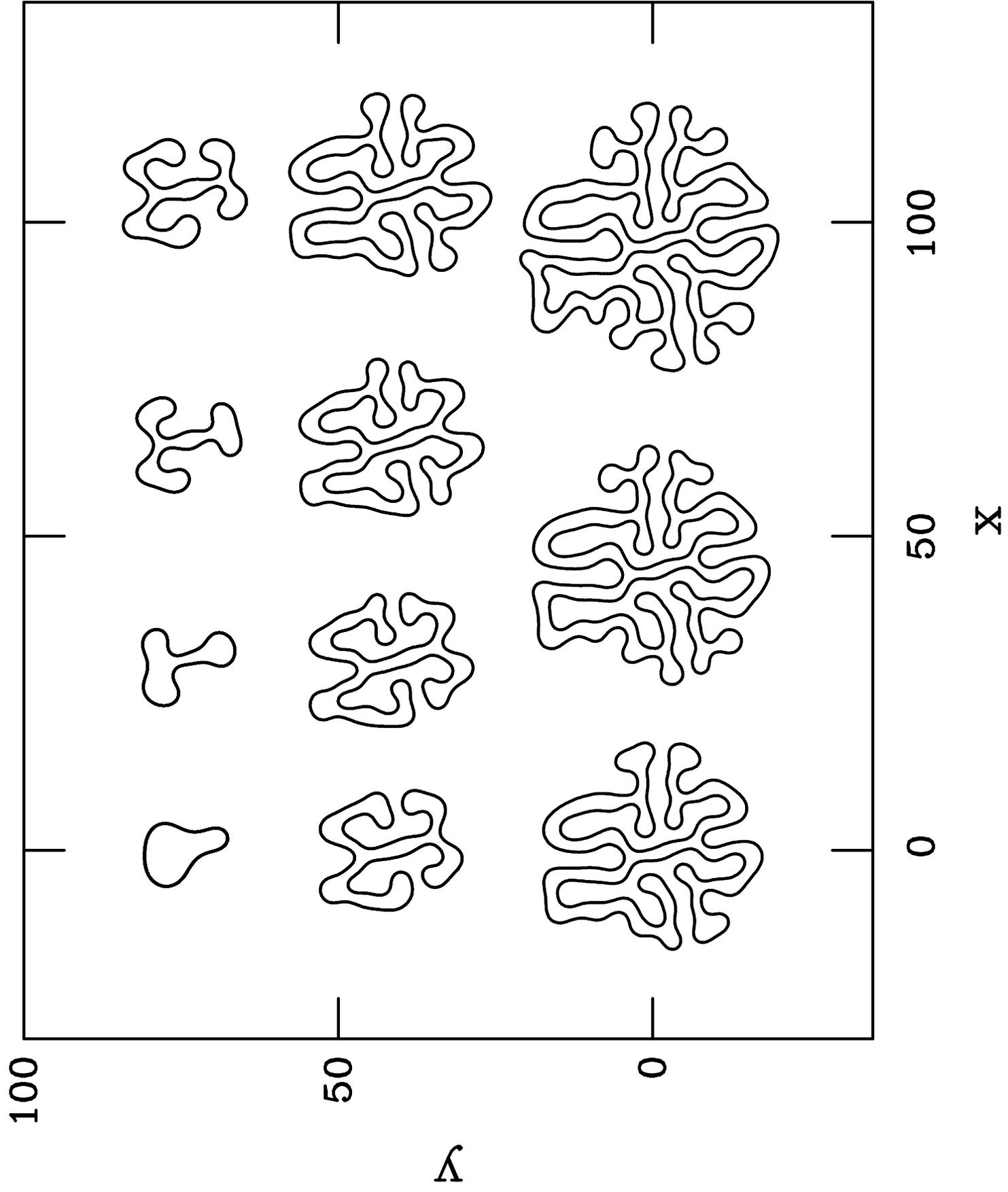
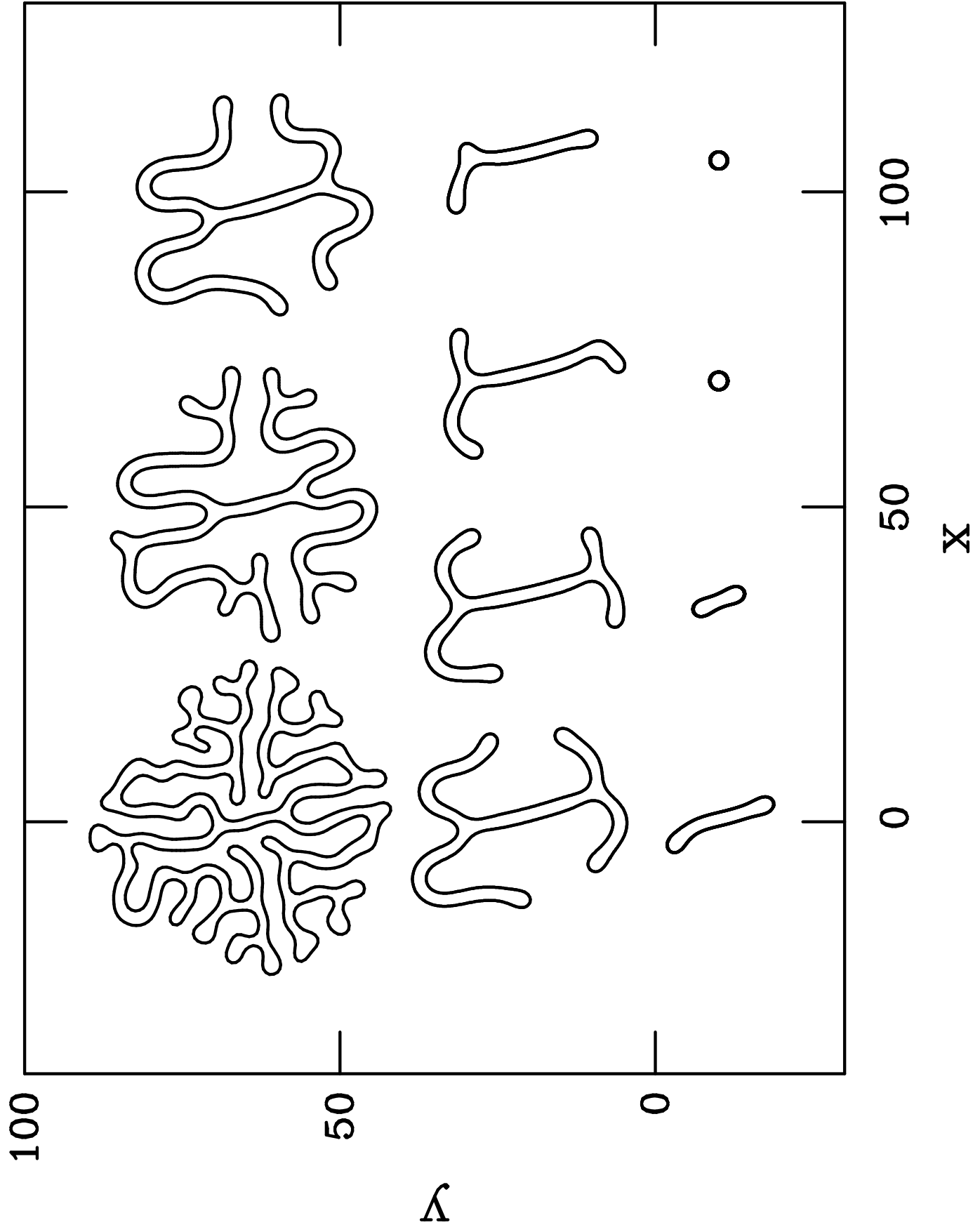
\end{document}